%% file: main.tex
\documentclass{article}

\usepackage{nexerrar1}
\usepackage{graphicx}
\usepackage{hyperref}
\usepackage{etoc}
\usepackage{amsmath}
\usepackage{floatrow}
\usepackage{xcolor}
\usepackage{caption}
\usepackage{tabularray}
\UseTblrLibrary{booktabs}
\usepackage{tocloft}

\title{A chemical language model for reticular materials design}
\author{%
  Dhruv Menon\textsuperscript{1}, Vivek Singh\textsuperscript{2,3,4},
  Xu Chen\textsuperscript{1}, Mohammad Reza Alizadeh Kiapi\textsuperscript{1},
  Ivan Zyuzin\textsuperscript{1}, Hamish W. Macleod\textsuperscript{1}, Nakul Rampal\textsuperscript{2,3,4}, William Shepard\textsuperscript{5}, Omar M. Yaghi\textsuperscript{2,3,4} and David Fairen-Jimenez\textsuperscript{1,*}%
}

\DMAffiliation{\textsuperscript{1}Department of Chemical Engineering \& Biotechnology, University of Cambridge, Cambridge, UK \newline \textsuperscript{2}Department of Chemistry, University of California – Berkeley, Berkeley, CA, USA \newline  \textsuperscript{3}Bakar Institute of Digital Materials for the Planet, Berkeley, CA, USA \newline  \textsuperscript{4}KACST–UC Berkeley Center of Excellence for Nanomaterials for Clean Energy Applications, King Abdulaziz City for Science and Technology, Riyadh, Saudi Arabia \newline  \textsuperscript{5}Synchrotron SOLEIL-UR1, L’Orme des Merisiers, Départementale 128, 91190 Saint-Aubin, France}
\DMDate{March 19, 2026}
\DMCorrespondingAuthor{df334@cam.ac.uk}
\DMCopyright{2026 The Authors. All rights reserved}
\DMHeaderImage{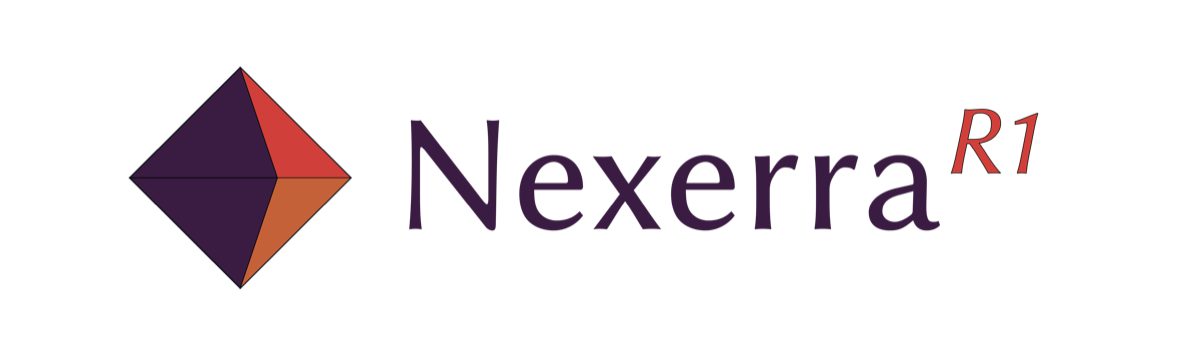}
\DMAbstract{%
Reticular chemistry has enabled the synthesis of tens of thousands of metal–organic frameworks (MOFs), yet the discovery of new materials still relies largely on intuition-driven linker design and iterative experimentation. As a result, researchers explore only a small fraction of the vast chemical space accessible to reticular materials, limiting the systematic discovery of frameworks with targeted properties. Here, we introduce Nexerra$^{R1}$, a building-block chemical language model that enables inverse design in reticular chemistry through the targeted generation of organic linkers. Rather than generating complete frameworks directly, Nexerra$^{R1}$ operates at the level of molecular building blocks, preserving the modular logic that underpins reticular synthesis. The model supports both unconstrained generation of low-connectivity linkers and scaffold-constrained design of symmetric multidentate motifs compatible with predefined nodes and topologies. We further combine linker generation with flow-guided distributional targeting to steer the generative process toward application-relevant objectives while maintaining chemical validity and assembly feasibility. The generated linkers are subsequently assembled into three-dimensional frameworks and structurally optimized to produce candidate materials compatible with experimental synthesis. Using Nexerra$^{R1}$, we validate this strategy by rediscovering known MOFs and by proposing the experimental synthesis of a previously unreported framework, CU-525, generated entirely \textit{in silico}. Together, these results establish a general inverse-design paradigm for reticular materials in which controllable chemical language modelling enables the direct translation from computational design to synthesizable frameworks.}

\begin{document}
\makedmtitle
\noindent \textbf{Public repository} : \href{https://github.com/fairen-group/nexerra-r1}{Nexerra$^{R1}$} $|$ \textbf{Model weights} : \href{https://doi.org/10.5281/zenodo.19100678}{Zenodo}
\section{Introduction}
\input{sections/intro.tex}
\section{Model Architecture}
\input{sections/architecture}
\section{Linker design and de novo MOF synthesis}
\input{sections/linker}
\section{Flow-guided design}
\input{sections/flow}
\section{Outlook}
\input{sections/discussions}
\section{Data Availability}
\input{sections/data}
\section{Code Availability}
\input{sections/code}
\section{Author Contributions}
\input{sections/author}
\section{Acknowledgements}
\input{sections/acknowledgements}
\section{Conflicts of Interests}
\input{sections/conflicts}


\clearpage
\begingroup
\title{Supplementary Information}

\author{Dhruv Menon\textsuperscript{1}, Vivek Singh\textsuperscript{2,3,4},
  Xu Chen\textsuperscript{1}, Mohammad Reza Alizadeh Kiapi\textsuperscript{1},
  Ivan Zyuzin\textsuperscript{1}, Hamish W. Macleod\textsuperscript{1}, Nakul Rampal\textsuperscript{2,3,4}, William Shepard\textsuperscript{5}, Omar M. Yaghi\textsuperscript{2,3,4} and David Fairen-Jimenez\textsuperscript{1,*}
  }

\DMAbstract{}
\makedmtitle

\setcounter{section}{0}
\setcounter{figure}{0}
\setcounter{table}{0}

\renewcommand{\thesection}{S\arabic{section}}
\renewcommand{\thefigure}{S\arabic{figure}}
\renewcommand{\thetable}{S\arabic{table}}

\thispagestyle{empty}
\clearpage
\renewcommand{\thesection}{\Alph{section}}
\numberwithin{figure}{section}
\renewcommand{\thefigure}{\Alph{section}\arabic{figure}}
\section{Supplementary Section A. Edge-transitive nets and compatible nodes}
\label{sec:A}
\input{sections/A}

\section{Supplementary Section B. Nexerra$^{R1}$ architecture}
\input{sections/B}

\section{Supplementary Section C. Dataset curation and generation}
\input{sections/C}

\section{Supplementary Section D. Model training \& evaluation}
\input{sections/D}

\section{Supplementary Section E. Post-processing of linkers after generation}
\input{sections/E}
\section{Supplementary Section F. Reward functions}
\input{sections/F}

\section{Supplementary Section G. Design modes}
\input{sections/G}

\section{Supplementary Section H. Experimental section}
\input{sections/H}
\section{Supplementary Section I. Flow model}
\input{sections/I}
\section{Supplementary Section J. Computational details}
\input{sections/J}
\endgroup
\clearpage
\bibliographystyle{plainnat}
\bibliography{references}
\end{document}

%% file: sections/intro.tex
The geometry-guided design and assembly of periodically extended structures through reticular chemistry has transformed the discovery of porous materials, enabling breakthroughs in areas such as carbon capture \citep{zhou2024carbon, chen2025flexibility, lin2021scalable} and water harvesting \citep{song2023mof}. Over the past three decades, researchers have discovered most metal-organic frameworks (MOFs) through chemical-intuition-driven experimentation. While this approach has delivered remarkable materials with exceptional properties, progress has remained largely empirical and slow, which has limited the systematic exploration of the vast chemical space accessible to MOFs and limited their rapid translation into industrial applications \citep{chakraborty2025make, asgari2025structuring, wright2025transitioning}. Computational methods have begun to address this challenge. Classical molecular simulations and machine-learning (ML) approaches now allow researchers to screen large libraries of materials \textit{in silico}, expanding the search space and reducing experimental effort \citep{jablonka2020big, moghadam2024progress}. Despite these advances, current strategies remain fundamentally constrained. High-throughput screening still explores only a tiny fraction of the enormous design space of MOFs and relies entirely on known structures or, at best, combinatorially assembled hypothetical frameworks. As a result, current approaches struggle to guide the discovery of genuinely new materials designed from the outset to exhibit target properties.\newline
\indent \indent \indent In this context, generative modelling offers a promising route towards the inverse design of MOFs with targeted properties. These approaches learn the underlying statistical structure of chemical datasets and use this knowledge to generate new candidates that satisfy predefined objectives. Several generative approaches have recently emerged, including chemical language models, score-based models, and denoising diffusion probabilistic models (DDPMs), which have shown considerable success across molecular design problems such as small-molecule generation \citep{elijovsius2025zero}, drug discovery \citep{krishnan2025generative}, \textit{de novo} protein design \citep{bennett2026atomically}, reaction mechanism prediction \citep{joung2025electron} and inorganic materials design \citep{zeni2025generative}. Among these approaches, chemical language models (CLMs) have attracted particular attention. These models treat molecular representations as sequences and learn to predict the probability of the next token given the preceding context \citep{zhao2023survey, skinnider2021chemical}. By learning the statistical patterns embedded in larger chemical datasets, CLMs can generate entirely new molecules consistent with the training distribution \citep{skinnider2021chemical}. When trained on massive datasets, these architectures scale into large language models (LLMs), capable of capturing increasingly complex relationships \citep{zhao2023survey}. Other generative strategies have also proven powerful. Diffusion-based approaches, such as DDPMs, iteratively denoise data corrupted with noise, generating highly diverse outputs of interest \citep{igashov2024equivariant}. Score-based models provide a closely related formulation in which neural networks estimate gradients of the logarithmic probability density (known as the score) of noise-corrupted data \citep{lee2023score}. Despite this rapid progress, applying generative modelling to reticular chemistry remains challenging. Many existing approaches struggle to ensure synthetic feasibility, adapt to the modular design principles of reticular materials, or generate structures that remain compatible with experimentally accessible building blocks and topologies. As a result, significant opportunities remain to develop generative strategies that directly address the unique constraints of reticular synthesis and enable the targeted discovery of new MOFs \citep{cleeton2025inverse, yao2021inverse, park2025multi, zheng2025large}.\newline
\indent \indent \indent We reason that decoupling the design process at the building-block level (\textbf{Figure 1a}) would provide stronger control over generative workflows and enable more targeted MOF discovery. This approach mimics experimental practice, where synthetic chemists rationally select or design molecular building blocks (MBBs) with pre-specified geometries to assemble desired nets \citep{jiang2021reticular}. Here, we introduce Nexerra$^{R1}$, an unsupervised CLM that generates diverse, synthesizable, organic linkers that can be fine-tuned towards a wide range of application-specific properties of interest, enabling inverse MOF design. Nexerra$^{R1}$ supports both the unconstrained design of bidentate and tridentate linkers, and the scaffold- or functional motif-constrained design of symmetric, multidentate linkers (e.g., porphyrin- and pyrene-derived linkers). Under the formulation of flow-matching \citep{lipman2022flow, tong2023improving} – a generalization of the diffusion process – in an optimal transport (OT)-setting \citep{tong2023improving},  Nexerra$^{R1}$ can be steered towards target molecular distributions relevant to specific applications. By combining theory-informed reward functions with design templates, Nexerra$^{R1}$ can guide the design of high-performing candidates that are compatible with experimental synthesis.

%% file: sections/architecture.tex
Designing new MOFs often begins at the level of molecular building blocks \citep{jiang2021reticular}. In reticular chemistry, frameworks arise from the assembly of inorganic nodes and organic linkers into periodic nets whose geometry determines the resulting structure (\textbf{Figure 1a}). To make the design space tractable and ensure reliable targets, we focus on edge-transitive nets (i.e., nets with a single type of non-intersecting edge) and zero-periodic metal-cluster building blocks (MBBs), which are experimentally robust, yet span diverse coordination geometries (see \textbf{Supplementary Information Section A} for representative nets and nodes) \citep{jiang2021reticular, guillerm2025elementary}. Linkers offer a vast design space, where subtle structural changes can induce major effects on MOF structure and function. Nexerra$^{R1}$ uses a textual representation of linkers and learns a continuous latent space for inverse design using a $\beta$-variational autoencoder ($\beta$-VAE) \citep{burgess2018understanding} as a probabilistic generative model (for model architecture, see \textbf{Supplementary Information Section B.2}). To enable this representation, we encode linkers using SELFIES, a molecular string language with formal grammar that guarantees chemical validity (\textbf{Supplementary Information Section B.1}) \citep{krenn2022selfies}.\newline 
\begin{figure}[b!]
\centering
\includegraphics[width = \textwidth]{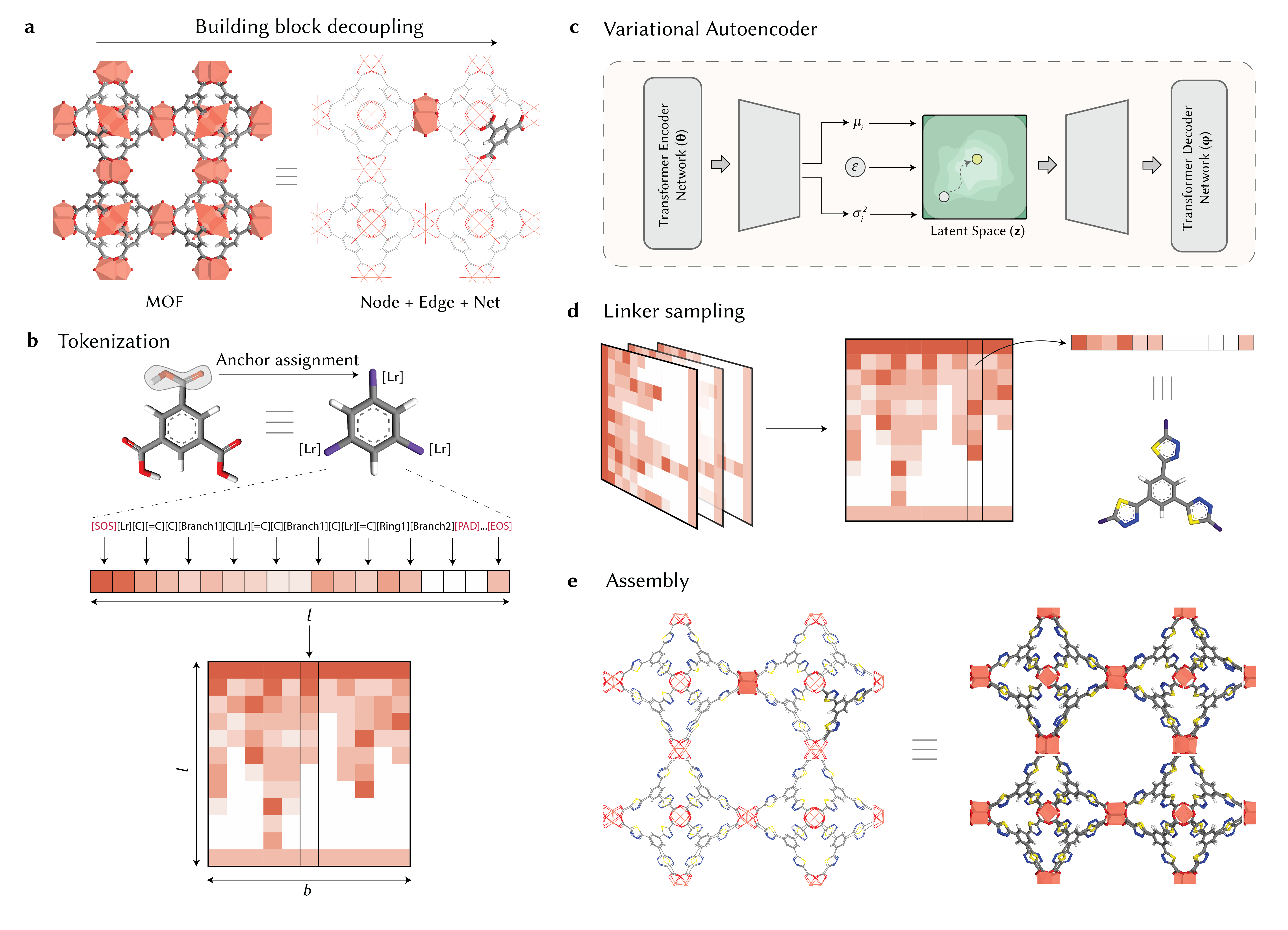}
\caption{\textbf{Inverse reticular chemistry with Nexerra$^{R1}$. a.} MOF structures are represented as nets made of nodes and edges. As a first step, we decouple the design process at the building block level. \textbf{b.} Linkers are encoded as molecular strings (SELFIES) and tokenized into a sequence of symbols. Each token acts as a structural unit that is processed by the model. \textbf{c.} The VAE acts as a bottleneck wherein the encoder defines an approximate posterior over a low-dimensional, continuous latent variable (z), while the decoder reconstructs the original sequence from z. \textbf{d.} Nucleus sampling is used to generate new linkers from the decoder output. \textbf{e.} Subject to the synthetic feasibility of the designed linker, an attempt is made to assemble it into the target net. (A detailed theoretical description of the model is available in \textbf{Supplementary Information Section B}. \textbf{Supplementary Information Figure B.1} describes the encoder and decoder neural networks. Colour code: Polyhedron - Cu$_2$($\mu$ - COO)$_4$; Gray - C; Red - O; Blue - N; Yellow - S; White - H; Purple - metal-linker coordination motif).}
\end{figure}
\indent \indent \indent Each SELFIES string is then tokenized, with each token acting as a structural unit processed by a neural network (\textbf{Figure 1b}). The encoder network approximates a posterior $q_\phi\left(z\middle| x\right)$ over a low-dimensional, continuous latent variable $z$; the decoder specifies a likelihood $p_\theta\left(x\middle| z\right)$ that reconstructs the original sequence from $z$ (\textbf{Figure 1c}). The model is trained by maximizing a $\beta$-weighted evidence lower bound, balancing accurate input reconstruction against latent distribution regularization (for mathematical formulation, see \textbf{Supplementary Information Section B.2}). This continuous latent space enables efficient sampling, smooth exploration and controlled perturbation of the linker chemical space for inverse design \citep{gomez2018automatic}.\newline 
\indent\indent\indent The encoder and decoder are implemented as stacks of transformer layers (outlined in \textbf{Supplementary Information Figure B.1}), transforming a set of vectors from one representation space to another while learning a rich internal representation \citep{lin2022survey}. Through the attention mechanism \citep{vaswani2017attention}, which assigns input-dependent weights to different elements of the sequence, capturing long-range dependencies, molecular strings are mapped to chemically meaningful embeddings that support generative design in the latent space (\textbf{Supplementary Information Section B.3}). Specifically, the encoder contextualizes the input token embeddings and compresses them into a latent representation through the self-attention mechanism, while the decoder projects this representation back into the input space using a combination of masked self-attention and cross-attention mechanisms (\textbf{Supplementary Information Section B.3}). The decoder outputs a probability distribution over the next token in the generated sequence, which we sample to generate new linkers (\textbf{Figure 1d}, \textbf{Supplementary Information Section B.4}). Following post-processing, where the placement of coordination moieties are optimised (optimisation algorithm described in \textbf{Supplementary Information Section E}), the linkers are subsequently assembled along with the node on a chosen net template to construct a new MOF (\textbf{Figure 1e}, \textbf{Supplementary Information Section E}).\newline
\indent\indent\indent Nexerra$^{R1}$ is trained on a corpus of approximately 900,000 linkers derived from existing databases \citep{yao2021inverse} and systematically functionalized to expand chemical diversity (for dataset curation, see \textbf{Supplementary Information Section C}). To maintain synthetic plausibility, we applied chemistry-based filters to remove linkers with high synthetic complexity, quantified by the synthetic complexity score (SCScore) \citep{coley2018scscore}. The resulting dataset spans a chemically rich, non-redundant linker space biased toward synthetically accessible chemotypes (\textbf{Supplementary Information Section C}). Nexerra$^{R1}$ achieves per-token reconstruction accuracy of 96.5\%, internal diversity > 0.925, and novelty > 96\%, consistent with a non-degenerate latent representation of MOF linker space (for training and evaluation, see \textbf{Supplementary Information Section D}). Distributions of 10,000 generated linkers against an unseen test set affirm that Nexerra$^{R1}$ reproduces marginal distributions of key descriptors (\textbf{Supplementary Figure D.2}). Under unconstrained sampling, Nexerra$^{R1}$ proposes diverse, multifunctional linkers with extended aromatic cores and dense substitution patterns consistent with the expressive capacity of the architecture (example linkers in \textbf{Supplementary Figure D.4}). However, unconstrained generation alone does not guarantee an efficient ‘hit rate’ of application-relevant candidates. We therefore employ templated, seed-guided design to focus sampling on promising regions of linker space and improve the efficiency of the inverse design process.\newline

%% file: sections/linker.tex
\noindent Unconstrained sampling draws latent vectors $z$ from the prior $p\left(z\right)$ and decodes them autoregressively into new linkers, enabling global exploration of the learned linker distribution. In practice, however, only a small fraction of such samples remain compatible with the geometric constraints imposed by a chosen node-net template and with application-driven design objectives. We therefore use Nexerra$^{R1}$ as a controllable generator via seeded inference: we encode a known linker ($x_0$) into a local latent distribution $q_\phi\left(z\middle| x_0\right)$, sample the neighbourhood of its embedding, and decode analogues that preserve the seed chemotype while exploring nearby functionalization and shape space. We couple this localized sampling to post-hoc ranking under user-defined reward functions (denoted $R$), retaining the highest-scoring candidates ($\mathcal{X}^\ast=TopK\left(x_i,R\right)$). \textbf{Supplementary Information Section F} shows the theoretical basis and formulation of these functions. Importantly, the generative model is not updated by the reward; $R$ is used solely to select designs.\newline
\indent\indent\indent We implement seeded inference in two complementary modes (\textbf{Figure 2a}, \textbf{Supplementary Information Section G}). In ‘direct design’, the entire seed ($x_0$) is allowed to vary. From the encoded neighbourhood of $x_0$, we generate a local library $\{x_k\}$, assemble each linker on the chosen reticular template, and retain the $TopK$ candidates under $R$. This mode is most effective for low-connectivity, i.e., highly-symmetric systems where the node geometry and linker connectivity largely determine topology. Using MOF-5 as a template and 1,4-benezenedicarboxylate (BDC) as a seed, we obtained 29 linker designs satisfying $R_{gas}$ > 0.75 $\times$ $R_{gas} (seed)$ – $R_{gas}$ being a function formulated for identifying exceptional linkers for gas storage (derived in \textbf{Supplementary Information Section F}). We could assemble nearly all candidates on the \textit{pcu} net, yielding multiple clean, force-field optimized frameworks. \textbf{Figure 2b} shows 6 such designed MOFs ($\phi_1\ -\ \phi_6$) after cleaning and optimization; \textbf{Supplementary Information Section G.1} shows the direct design approach applied to other classes of bidentate and tridentate linkers.\newline
\begin{figure}[t!]
\centering
\includegraphics[width = \textwidth]{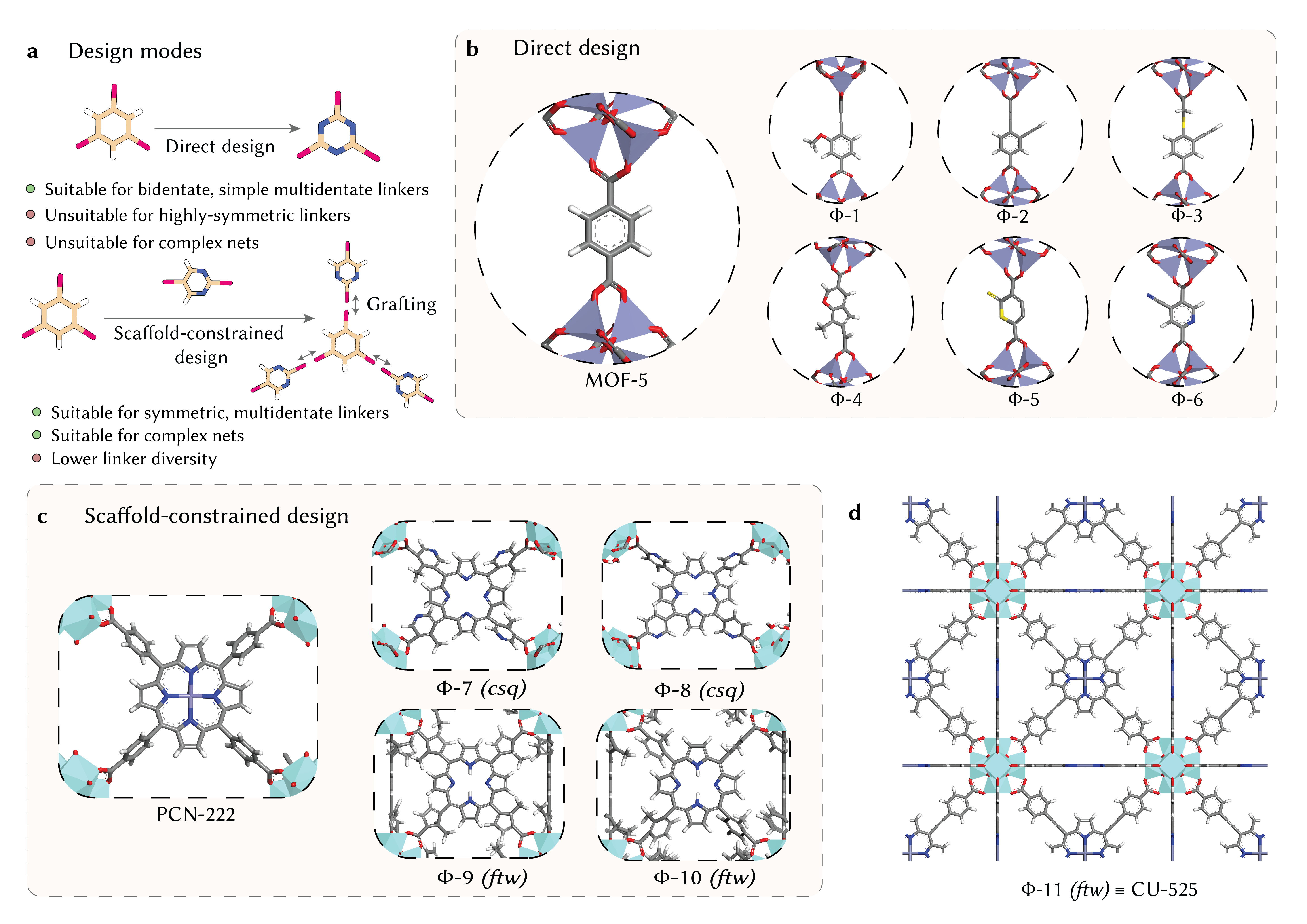}
\caption{\textbf{\textbf{Linker design.}} \textbf{a.} Nexerra$^{R1}$ enables inference in two modes: direct design and scaffold-constrained design. In direct design, Nexerra$^{R1}$ fully maps the seed to new linkers while in scaffold-constrained design, the scaffold is fixed, and the model generates new arms that are symmetrically grafted to the core. \textbf{b.} Using MOF-5 as a design template, Nexerra$^{R1}$ generates a library of linkers which are the filtered and scored by a reward function (R). $\phi_1\ -\ \phi_6$ are 6 representative MOFs that were successfully assembled with the Zr$_4$O node on the \textit{pcu} net. \textbf{c.} Using PCN-222 as a design template, $\phi_7\ -\ \phi_{10}$ are 4 representative MOFs that were assembled with the Zr$_6$-oxo node. Since several topologies can be targeted (depending on geometric and topological constraints), we were able to successfully assemble candidates into the 8-connected \textit{csq} net and the 12-connected \textit{ftw} net. \textbf{d.} $\phi_{11}$ assembled on a \textit{ftw} net was experimentally synthesized as CU-525. (Colour code: Polyhedron (Purple) – Zn$_4$O(COO)$_6$; Polyhedron (Blue) – Zr$_6$($\mu_3$-O)$_4$($\mu_3$-OH)$_4$(COO)$_8$(OH/H$_2$O)$_4$; Gray - C; Red - O; Blue - N; Yellow - S; White - H)}
\end{figure}
\indent\indent\indent For high-connectivity frameworks, direct design becomes unreliable because small perturbations in the latent space can distort rigid, highly symmetric scaffolds that are essential for preserving topology. We, therefore, introduce a second design strategy, ‘scaffold-constrained design’ (\textbf{Figure 2a}), in which the linker is decomposed as $x=\left(s,a\right)$, where $s$ represents a fixed core scaffold, and $a$ represents variable arms. During generation, we preserve the scaffold and sample only the arms, allowing Nexerra$^{R1}$ to explore functionalization and pore-environment chemistry while preserving the connectivity and geometry required for framework assembly. We apply this strategy to  porphyrinic Zr$_6$-oxo systems using PCN-222 as a template. Starting from the tetrakis(4-carboxyphenyl)porphyrin (TCPP) linker, Nexerra$^{R1}$ diversifies linker chemistry while retaining the porphyrin core, enabling targeted modulation of the pore-environment without violating topology-critical geometry (\textbf{Figure 2c}). Among the top-ranked candidates (filtered using $R_{gas}$), several could be assembled and optimized on \textit{csq} and \textit{ftw} nets. \textbf{Figure 2c} shows four such designed MOFs ($\phi_7-\ \phi_{10}$); \textbf{Supplementary Information Section G.2} shows this approach applied to other linkers. Notably, the Hf$_6$-oxo analogue of one such \textit{ftw} design was synthesised as CU-525 (CU = Cambridge University), a variant of MOF-525 (\textbf{Figure 2d}). This result demonstrates that scaffold-constrained generative design can yield synthetically accessible, highly porous frameworks whose internal chemistry can be tailored for function. Single crystal X-ray diffraction (SCXRD) data were collected and the crystal structure of CU-525 was successfully determined on the micro-focus beamline \textit{PROXIMA 2A} at Synchrotron SOLEIL (for experimental details, see \textbf{Supplementary Information Section H}). Although this seeded approach concentrates sampling relative to unconstrained generation, it still relies on random exploration followed by ranking. The learned prior reflects the empirical distribution of the training data rather than the structural requirements of a given application. Consequently, sampling from this prior – even when guided by seeds – remains inefficient when targeting linkers with specific geometric or physicochemical attributes.\newline 

%% file: sections/flow.tex
\noindent To address this sampling inefficiency, we developed a flow model – Nexerra$^{R1}$-Flow – that implements a continuous normalizing flow in the latent space of Nexerra$^{R1}$ (\textbf{Figure 3a}) \citep{tong2023improving}. The flow defines a time-dependent, invertible transformation by learning a vector field $v_\theta\left(z,t;c\right)$ that transports latent samples ($z$) from a base distribution $p_0$ to a target distribution $q_1$; conditioning is introduced using a scalar variable $c$ \citep{lipman2022flow, tong2023improving}. Nexerra$^{R1}$-Flow is trained using optimal-transport conditional flow matching (OT-CFM); an optimal transport coupling ($\pi$) is computed between minibatches from $p_0$ and $q_1$, and intermediate points $x_t$ together with conditional velocities $v_t$ are sampled along the corresponding displacement interpolation path. A neural ordinary differential equation (ODE) is then trained to match these velocities, allowing probability mass to be redistributed toward the target regime without modifying the underlying CLM (for mathematical formulation, see \textbf{Supplementary Information Section I.1}). This flow formulation is inherently reward-agnostic; $q_1$ can be biased toward arbitrary scalar objectives. The difficulty of this transport, however, reflects the intrinsic geometry of the objective in the latent space. Fragmented or non-smooth reward landscapes produce more complex transport dynamics – making the training process more difficult.\newline
\begin{figure}[h!]
\centering
\includegraphics[width = \textwidth]{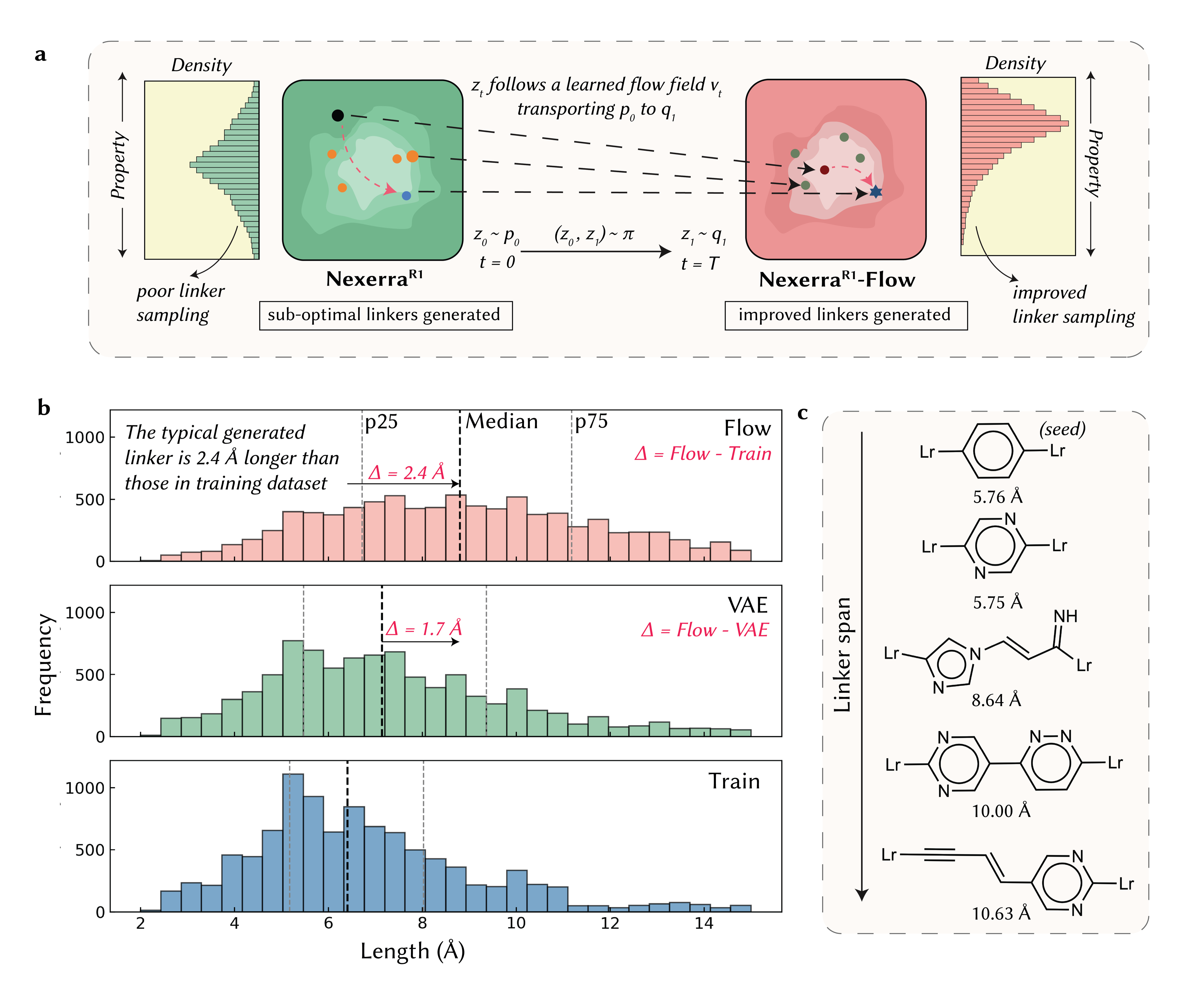}
\caption{\textbf{Flow-guided design. a.} Nexerra$^{R1}$-Flow learns a vector field $v_\theta\left(z,t;c\right)$ that transports latent samples from a base distribution $p_0$ to a target distribution $q_1$ under the formulation of optimal-transport conditional flow matching (a detailed theoretical description of the model is available in \textbf{Supplementary Information Section I.1}). \textbf{b.} Linker span distributions of molecules sampled from the training dataset (n = 10,000), Nexerra$^{R1}$ (n = 10,000) and Nexerra$^{R1}$-Flow (n = 10,000). The typical generated linker from Nexerra$^{R1}$-Flow is 2.4 Å longer compared to the training dataset and 1.7 Å longer compared to Nexerra$^{R1}$. \textbf{c.} Using 1,4-benzenedicarboxylic acid as the seed, Nexerra$^{R1}$-Flow generates several optimal linkers with a longer span; the best-ranked linker has a span of 10.63 Å which is 1.8$\times$ the span of the seed. The corresponding generated structures are presented in \textbf{Supplementary Information Section I.4}.}
\end{figure}
\indent\indent\indent The linker length (defined by us as the maximum anchor – anchor span) presents a physically interpretable descriptor. Increasing the span helps maximise accessible pore volume, lending itself to applications such as improved gravimetric gas storage and heterogeneous catalysis \citep{chen2020balancing, chen2025topology, le2026oxidative}. Starting from an empirical latent prior $p_0$, we train Nexerra$^{R1}$-Flow toward a target distribution biased towards longer spans by progressively increasing percentile thresholds. Specifically, over 200,000 steps, the target region was annealed from the 70$^{th}$ to the 95$^{th}$ percentile of the anchor – anchor span distribution (for training and evaluation, see \textbf{Supplementary Information Section I.3}). This progressively biases generation toward longer linker spans. \textbf{Figure 3b} shows span distributions of the linkers sampled from the training dataset, Nexerra$^{R1}$ and Nexerra$^{R1}$-Flow respectively. While Nexerra$^{R1}$ reproduces the marginal distributions of the training dataset (KL(Train $||$ VAE) $\approx$ 0.10), there is a marked shift in the distribution generated by the flow model (KL(Train $||$ Flow) $\approx$ 0.42). Notably, the median length sampled from Nexerra$^{R1}$-Flow is right-shifted by 2.4 Å compared to the training dataset and by 1.7 Å compared to Nexerra$^{R1}$. There is a large population of generated linkers that exceed 10 Å in span. As an illustrative example, using the 1,4-benzenedicarboxylic acid linker (BDC) as a seed, Nexerra$^{R1}$-Flow generated several linkers that outperform it pursuant to a $R$ that prioritises length. \textbf{Figure 3c} shows several optimal candidates generated by Nexerra$^{R1}$-Flow subject to strict geometrical filters to enable assembly into the \textit{fcu} net; \textbf{Supplementary Figure I.3} shows the generated structures post-optimization. Notably, Nexerra$^{R1}$-Flow was able to scale the linker span from 5.76 Å for the seed to over 10.6 Å for the best generated candidate.\newline 
\begin{figure}[h!]
\centering
\includegraphics[width = 0.8\textwidth]{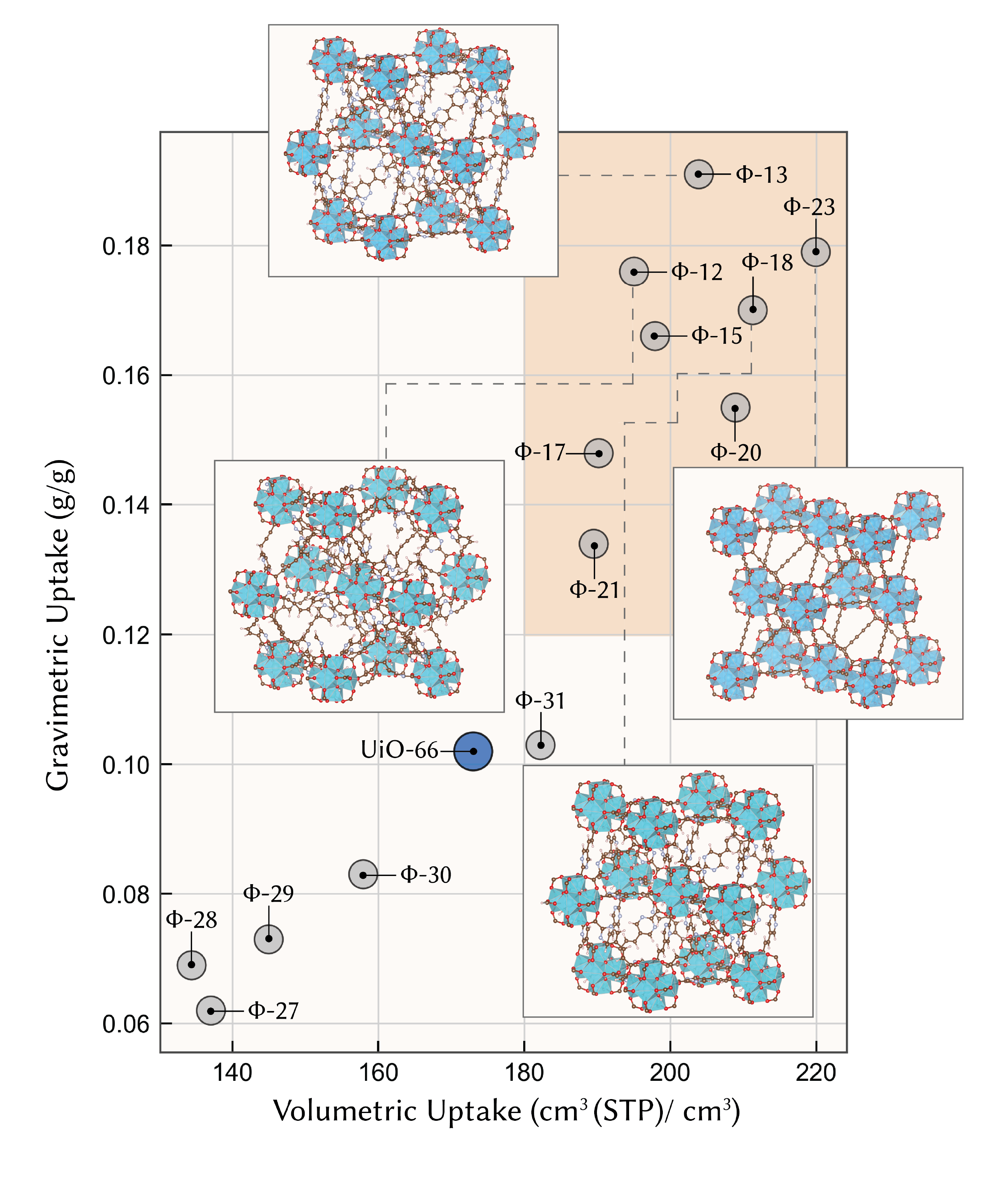}
\caption{\textbf{Flow-guided inverse design of UiO-66 analogues for methane storage.} Gravimetric versus volumetric methane uptake at 298 K and 80 bar calculated using GCMC simulations for \textit{fcu} frameworks assembled from Nexerra$^{R1}$-Flow designed linkers (BDC seed). Several candidates simultaneously exceed UiO-66 in both volumetric and gravimetric uptake. The distributional steering achived by Nexerra$^{R1}$-Flow enables concurrent improvements in storage performance as shown in the darker shaded region. Insets show representative high-performing optimized structures. (Colour code: Polyhedron – Zr$_6$($\mu_3$-O)$_4$($\mu_3$-OH)$_4$(COO)$_{12}$; Gray - C; Red - O; Blue - N; Yellow - S; White - H)}
\end{figure}
\indent \indent \indent To evaluate whether flow-guided latent transport translates into application-relevant performance gains, we simulated methane adsorption at 298 K and 80 bar using grand canonical Monte Carlo (GCMC) simulations on the optimized frameworks assembled on the \textit{fcu} net from the designed linkers discussed above (for computational details, see \textbf{Supplementary Information Section J}). \textbf{Figure 4} presents the gravimetric versus volumetric methane uptake landscape of the generated frameworks relative to UiO-66. The structures were able to achieve a systematic displacement of uptakes towards the upper-right region of the performance space; in other words, we were able to achieve a concurrent increase in both volumetric and gravimetric uptake – two typically competing metrics within a fixed topology \citep{chen2020balancing, chen2025topology}. While UiO-66 occupies a relatively moderate capacity (0.102 g/g gravimetric uptake; 173 cm$^3$ (STP)/cm$^3$ volumetric uptake), several generated MOFs significantly outperform the material. The two best-performing candidates – $\phi_{13}$ and $\phi_{23}$ – achieve gravimetric uptakes of 0.191 g/g and 0.179 g/g and volumetric uptakes of 203 cm$^3$ (STP)/cm$^3$ and 219 cm$^3$ (STP)/cm$^3$, respectively. These represent up to an 87\% increase in gravimetric and 27\% increase in volumetric uptake relative to UiO-66. These results establish that latent-space flow transport can efficiently bias reticular design toward high-performance regions of the materials space without altering topology.\newline
\indent \indent \indent More broadly, the platform is not restricted to gas storage. Since the design objective only enters through the reward function $\textit{R}$, alternate application-specific targets can be incorporated without modifying the underlying generative prior. As an illustrative example, building on our recent work on the computational design of MOFs for drug delivery \citep{menon2025guiding, melle2025rational}, in \textbf{Supplementary Information Section F.2} we develop a modular reward function $\textit{R}_{bio}$ for the design of biocompatible, high-performance linkers. This highlights how the same platform can be adapted to steer generative reticular chemistry to diverse functional regimes. Future work will couple high-ranked designs with experimental synthesis and characterisation for a direct assessment of the predictive power of flow-guided latent transport – ultimately, enabling accelerated reticular materials discovery.

%% file: sections/discussions.tex
We have introduced a targeted, controllable approach towards inverse reticular chemistry in which chemical language modeling is coupled to latent-space transport to bias generation toward functional regimes. The developed formulation is inherently reward-agnostic and extensible to arbitrary scalar objectives, including multi-objective applications potentially spanning adsorption selectivity, mechanical stability, catalytic turnover or drug delivery. A central implication of our work is the decoupling of representation learning from objective steering. By maintaining a fixed generative prior over the linker space and learning an optimal-transport map in latent space, we enable targeted redistribution of probability mass without the need for retraining. This allows application-specific design while preserving the structural grammar of reticular assembly. As such, the approach provides a general strategy for navigating constrained chemical design spaces where topology, symmetry and connectivity impose non-trivial geometric priors. We foresee that future developments will expand this methodology to joint linker–node optimization enabling the exploration of heterogeneous metal environments and mixed-node systems. Coupling latent transport to topological selection may further permit controlled navigation across nets, enabling simultaneous optimization of geometry and composition rather than operating within a fixed reticular scaffold. More broadly, as generative models become increasingly expressive and property evaluations increasingly automated, such approaches will provide a scalable route toward autonomous design of crystalline materials \citep{rong2025algorithmic}. We anticipate that platforms such as Nexerra$^{R1}$ will accelerate the transition from intuition-driven synthesis to programmable reticular engineering, where application-specific materials are generated, assembled and validated within a unified computational-experimental pipeline.

%% file: sections/data.tex
The crystallography data in this work have been deposited in the Cambridge Crystallographic Data Centre (CCDC) under accession number CCDC: 2514566 (CU-525). This data can be obtained free of charge from CCDC via \href{https://www.ccdc.cam.ac.uk/data_request/cif}{www.ccdc.cam.ac.uk/data\_request/cif}.

%% file: sections/code.tex
The code for Nexerra$^{R1}$ is available at \href{https://github.com/fairen-group/nexerra-r1}{https://github.com/fairen-group/nexerra-r1} under an MIT license. This repository will be under continuous development. A stable version (including model weights) will be released upon publication. 

%% file: sections/author.tex
D.M, O.M.Y and D.F-J conceived the idea. D.M led the model development with significant input from V.S, M.R.A.K and N.R. X.C experimentally synthesized CU-525. X.C. and W.S. collected SCXRD data and determined the X-ray crystal structure of CU-525. D.M and M.R.A.K performed the molecular simulations. D.M, H.W.M and I.Z formulated the reward functions. O.M.Y and D.F-J supervised the work. D.M wrote the first draft of the manuscript. All authors contributed to the preparation of the manuscript. 

%% file: sections/acknowledgements.tex
This work was supported by the Winton Cambridge – Berkeley Exchange Fellowship awarded to D.M. D.M acknowledges NanoDTC Cambridge – ESPSRC EP/S022953/1. M.R.A.K acknowledges support from the Cambridge Trust Scholarship and the Trinity Henry-Barlow Scholarship. H.W.M. acknowledges the Harding Distinguished Postgraduate Scholarship Programme. The authors acknowledge the allocation of beamtime at Synchrotron SOLEIL and the help of the PROXIMA 2A staff in performing SCXRD experiments.

%% file: sections/conflicts.tex
There are no conflicts to declare.

%% file: sections/A.tex
MOF structures can be described as periodic graphs (or nets) with coordination between nodes (vertex) and spacers (edges) \citep{guillerm2025elementary, jiang2021reticular, li2014topological}. In such a representation, framework assembly involves embedding this topological graph into Euclidean space under strict geometric and chemical constraints. Not all nets therefore, correspond to achievable design targets due to topological symmetry and connectivity heterogeneity – which, in turn, complicates algorithmic assembly and synthesizability.  We thus, focus on nets that are edge-transitive, i.e., edges which are symmetry-equivalent; each edge has identical local coordination and geometric identity \citep{jiang2021reticular}. These nets offer several advantages for inverse design. First, they require only a single type of linker, eliminating the combinatorial complexity of multivariate systems. Second, the geometric constraints are uniform, i.e., connectivity, span and orientation are consistent. Third, MOFs based on such nets are easier to experimentally synthesize. Notably, this constraint does not come at the expense of diversity; there are theoretically infinite structures possible on edge-transitive nets, forming a vast chemical space. \newline 
\indent \indent \indent A second constraint we impose is the use of zero-periodic metal-cluster building blocks (MBBs). These MBBs (such as Zr$_6$-oxo, Cu$_2$-paddlewheel, Zn$_4$O) have well-defined topological coordination behavior, making algorithmic assembly tractable. Such MBBs have led to the development of several chemically stable MOFs with well-defined properties \citep{jiang2021reticular, guillerm2025elementary}. In contrast, supramolecular building blocks (SBBs) \citep{guillerm2014supermolecular}, extended metal-chains \citep{jiang2024multivariate} or merged-net approaches \citep{jiang2024merged} introduce multiple additional degrees of freedom which drastically complicate algorithmic assembly and thus, inverse design. For detailed discussions on edge-transitive nets, MBBs and principles involving targeting such nets, we refer readers to some excellent reviews from Mohammed Eddaoudi’s group \citep{guillerm2025elementary, jiang2021reticular, guillerm2014supermolecular}.\newline 
\indent \indent \indent The successful assembly of nodes and designed edges on a target net would necessitate compatibility with geometric constraints. This would involve the connectivity of the linker being compatible with the target net, the span and orientation of the linker being within a reasonable range of the target values and the absence of steric clashes that would prevent coordination or lead to overlapping atoms. Certain nets such as \textit{pcu} and \textit{fcu} are much easier to target due to lower topological constraints, while nets with higher symmetry such as \textit{csq} are more difficult. Consequently, these ‘simpler’ nets will have higher success rates for assembly compared to the ‘difficult’ nets. Most nodes we explore here have coordination numbers ranging from 6 – 12 which include clusters such as Zr$_6$-oxo (seen in UiO-66, PCN-222) and Cu$_2$-paddlewheels (seen in HKUST-1).\newline 
\indent \indent \indent Our present work, therefore, does not address: (i) multi-linker or multi-edge nets \citep{jiang2024multivariate}; (ii) mixed-node systems; (iii) merged nets \citep{jiang2024merged}; (iv) periodic SBUs \citep{chen2025flexibility}; and (v) interpenetrated structures \citep{gupta2021control}. We foresee that future work will expand to the joint optimization of node geometry and improved algorithmic assembly, enabling the targeting of such complex nets.

%% file: sections/B.tex
\textbf{B.1. Sequence representation and SELFIES.} Nexerra$^{R1}$ is developed on molecular linkers represented as a sequence of tokens. The model is trained to predict the next token in a sequence given an initial set of tokens (\textbf{Supplementary Figure B.1.a-b}). Sequence representation is thus crucial to the development of a robust language model. We chose SELFIES (SELF-referencing embedded string) \citep{krenn2022selfies} as a string-based molecular representation for developing Nexerra$^{R1}$. The key advantage of SELFIES is that it is built on a formal grammar which guarantees chemical validity of generated sequences – which is a known problem of conventionally employed representations such as SMILES strings, especially on long sequences. For discussions on these formulated grammar rules and working principles of SELFIES, we refer readers to parent publications from Aspuru-Guzik and colleagues \citep{krenn2020self, krenn2022selfies}. \textbf{Supplementary Figure B.1.b} shows the SELFIES string of a representative linker.\newline 
\indent \indent \indent The SELFIES string (say $x$) is subsequently decomposed into a sequence of discrete tokens drawn from a finite (defined) vocabulary $V$, where each token corresponds to a structural unit. Let
\begin{equation}
    x = (x_1,. . .,x_L),\ \ \ x_t \in V
    \tag{B.1}
\end{equation}
denote the resulting tokenization of a SELFIES string. We append start-of-sequence ‘[SOS]’, end-of-sequence ‘[EOS]’ and padding ‘[PAD]’ symbols to enable autoregressive training. Each sequence is padded to a fixed maximum length (102 tokens) with the [PAD] symbol masked during training and inference.\newline 

\noindent \textbf{B.2. $\beta$-VAE.} Nexerra$^{R1}$ is formulated as a $\beta$-VAE with transformer-based encoder and decoder networks (\textbf{Supplementary Figure B.1.c-e}). While detailed mathematical motivations on the VAE and transformer networks remain outside the remit of present work - instead read \citep{kingma2013auto, vaswani2017attention} -, we formulate the working principles of the architecture. Given the token sequence $x$ from (Equation $B.1$), we introduce a continuous latent variable $z$ as follows,
\begin{equation}
   z \in \mathbb{R}^{d_z}; \ \ \ d_z = 128
    \tag{B.2}
\end{equation}
The probabilistic latent variable model defines,
\begin{equation}
   p_\theta (x, z) = p(z)p(x|z)
    \tag{B.3}
\end{equation}
with a Gaussian prior
\begin{equation}
   p(z) = \mathcal{N}(0, I)
    \tag{B.4}
\end{equation}
\noindent The transformer encoder maps these sequences to contextualized embeddings to produce a fixed-dimensional representation (\textbf{Supplementary Figure B.1.c}). From this, the approximate posterior is parameterized as follows,
\begin{equation}
    q_\phi (z|x) = \mathcal{N}\left(z; \mu_\phi(x), diag(\sigma_\phi^2(x))\right)
    \tag{B.5}
\end{equation}
\noindent The latent samples can be drawn from the posterior using the reparameterization trick,
\begin{equation}
    z = \mu_\phi(x) + \sigma_\phi(x) \odot \epsilon; \ \epsilon \sim \mathcal{N}(0, I)
    \tag{B.6}
\end{equation}
\noindent The transformer decoder (\textbf{Supplementary Figure B.1.e}) – conditioned on $z$ – defines an autoregressive likelihood,
\begin{equation}
    p_\theta(x|z) = \Pi_{t=1}^{L} p_\theta(x_t | x_{<t}, z)
    \tag{B.7}
\end{equation}

\noindent We linearly project $z$ from the latent space to the model space, and input it to the decoder via the cross-attention mechanism. The model parameters are optimized using a $\beta$-weighted evidence lower bound (ELBO),
\begin{equation}
    \mathcal{L} = \mathbb{E}_{q_{\phi}(z|x)}[log p_\theta(x|z)] - \beta D_{KL}\left(q_\phi(z | x)||p(z)\right)
    \tag{B.8}
\end{equation}
\noindent For reconstruction, we implemented a cross entropy loss with masked tokens. During training, to prevent posterior collapse, we gradually anneal the $\beta$-term starting from 0 during a warmup phase. Post-warmup, the $\beta$ is held at a fixed value for the rest of the training. Due to the powerful learning capabilities of the transformer neural network, we find $\beta$ = 0.01 to be an optimum value for stable training (for training curves, see \textbf{Supplementary Information Section D}).\newline  

\noindent \textbf{B.3. Transformer encoder and decoder.} Both the encoder and decoder are implemented as stacks of transformer layers (\textbf{Supplementary Figure B.1}). We use a standard implementation of the transformer architecture \citep{vaswani2017attention}. Given token embeddings, each encoder layer applies multi-head self-attention, 
\begin{equation}
   Attention(Q, K, V) = softmax\left(\frac{Q.K^T}{\sqrt{(d_k)}}\right)V
    \tag{B.9}
\end{equation}
followed by the conventional feedforward network and residual connections. Here, $K, Q, V$ are the key, query and values, respectively, and $d_k$ is the dimension of keys. In our context, the self-attention mechanism allows each token to attend to all others in the sequence – thereby capturing long-range structural dependencies within the linker. This is particularly important for MOF linkers which often have extended conjugation and coordination motifs where geometric relationships cannot easily be captured. The decoder too retains the attention-based working, but incorporates masked self-attention to preserve its autoregressive nature. In other words, we mask future tokens so that the decoder cannot ‘cheat’. We condition on the latent variable using cross-attention.\newline 
\indent\indent\indent We chose transformer networks for our architecture due to their ability to model long-range interactions and accommodate longer context windows without degradation. This is evidenced by high per-token reconstruction accuracies for a 102-token window. That being said, such architectures are not without their share of problems. The powerful learning capabilities require strong control over model hyperparameters to prevent memorization, while the high expressivity necessitates strong geometrical and chemical filters coupled with reward functions to generate desirable linkers.\newline  

\begin{figure}[!]
\centering
\includegraphics[width = 0.8\textwidth]{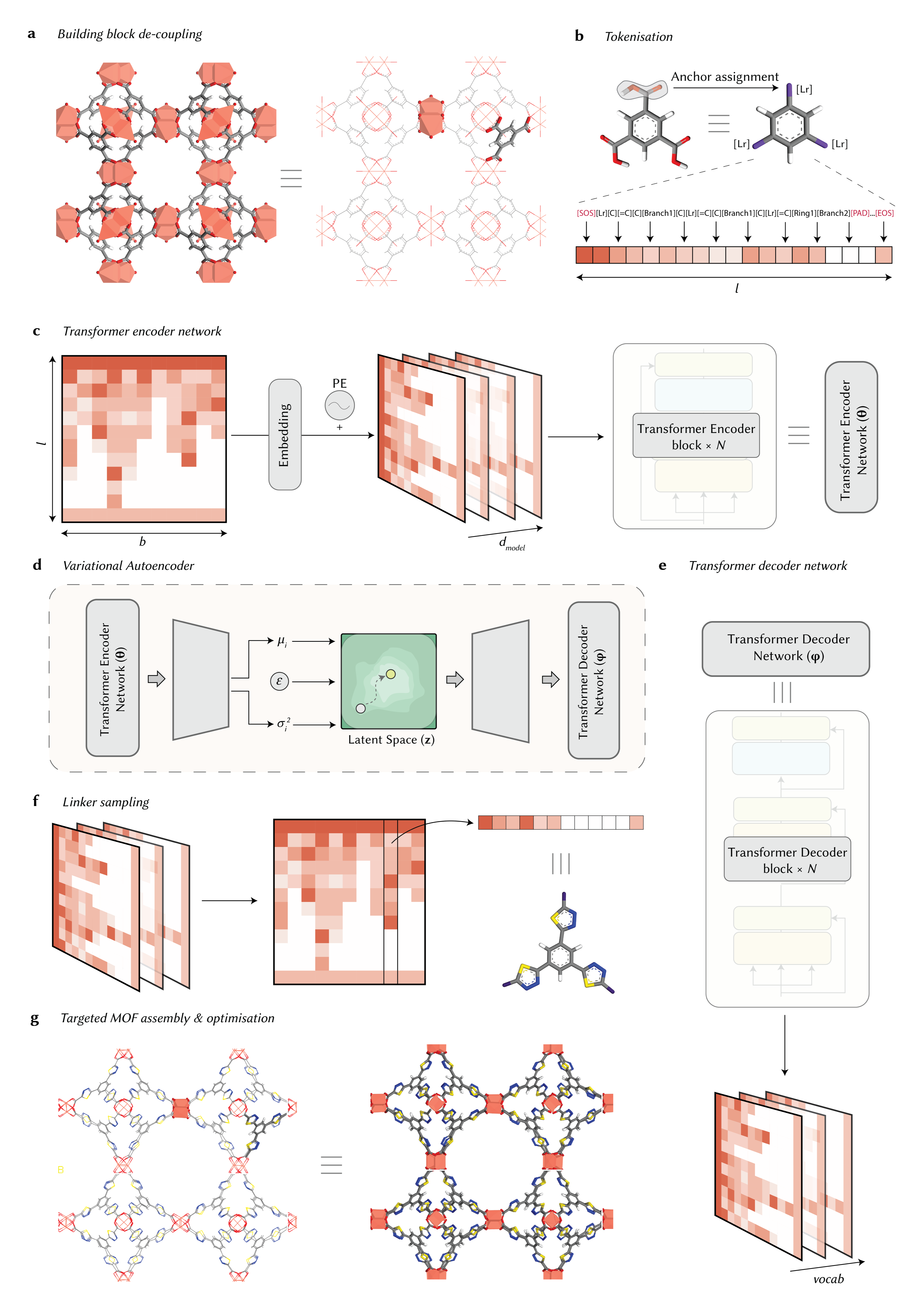}
\caption*{\textbf{Supplementary Figure B.1. Nexerra$^{R1}$ architecture. a.} MOF structures are represented as nodes and edges. As a first step, we decouple the design process at the building-block level. \textbf{b.} Linkers are encoded as molecular strings and tokenized into a sequence of symbols. \textbf{c.} The encoder network of the model is composed of a stack of transformer layers that map tokenized vectors in one representation space into a corresponding set of vectors in a new space, and by doing so, learn a rich internal representation. The $l \times b$ dataset ($l$ =  number of tokens; $b$ = number of samples) is first embedded and subsequently undergoes position encoding (PE) to capture sequence order. This embedded tensor is then passed to the transformer stack which applies the self-attention mechanism to capture long-range interactions. \textbf{d.} The VAE acts as a bottleneck wherein the encoder defines an approximate posterior over a low-dimensional continuous latent variable ($z$), while the decoder reconstructs the original sequence from $z$. \textbf{e} The decoder is again composed of a stack of transformer layers that projects the latent representation back into high-dimensional data and outputs logits of the vocabulary size (vocab). Here we implement masked self-attention to prevent the decoder from seeing future tokens during training. We condition it on the latent variable using cross-attention. \textbf{f.} Nucleus sampling is used to generate new linkers from the decoder output. \textbf{g.} Subject to chemical validity, an attempt is made to assemble the new linker into the target net (Color code: Polyhedron - Cu$_2$($\mu$-COO$_4$); Gray - C; Red - O; Blue - N; Yellow - S; White - H; Purple - [Lr])}
\end{figure}

\noindent \textbf{B.4. Inference.} At inference, during unconstrained sampling, new linkers are generated by sampling from the learned latent prior, $z \sim \mathcal{N}(0,I)$ followed by the autoregressive decoding of tokens described in Equation $B.7$. To control the extent of diversity (i.e., how similar or different samples be) we scale the generated logits by a temperature parameter $T( > 0)$. Lower values make the distribution sharper – i.e., less diversity – while higher values increase diversity. We get best results at $T = 0.8$, however sampling is flexible.\newline 
\indent \indent \indent To prevent sampling logits with extremely low probability (which lead to sub-optimal generated linkers) we perform nucleus (or top-p) sampling (\textbf{Supplementary Figure B.1.f}) \citep{holtzman2019curious}. This means that at $p = 0.9$, at each step, we only perform sampling on the subset of tokens whose cumulative probability mass exceeds 0.9. This eliminates sub-optimal tokens from consideration – improving inference stability. Inference continues until we hit an ‘[EOS]’ token or reach the maximum token limit of 102. Since SELFIES enforces validity, all generated molecules are syntactically and semantically valid. \newline 
\indent \indent \indent Following sampling, we implement a post-processing algorithm that optimizes the placement of coordination anchors – i.e., the [Lr] token. This algorithm is described in \textbf{Supplementary Information Section E}. Moreover, what we have described here is unconstrained sampling. A more efficient methodology involves seed-guided design (as introduced in the main manuscript). These design methods have been described in \textbf{Supplementary Information Section G}.

%% file: sections/C.tex
While there is a lack of consensus on how much data is required to train CLMs \citep{frey2023neural, skinnider2021chemical}, when it comes to specific domains such as MOF linkers, datasets spanning $10^5-10^6$ samples generally yield good results \citep{yao2021inverse}. Starting from the dataset reported by \cite{yao2021inverse}, we extracted $\sim$ 80,000 unique molecules (as SMILES). In order to expand the dataset, we implemented a randomized monovalent functionalization protocol to augment SMILES data. \textbf{Algorithm C.1} details the developed functionalization protocol. Briefly, for each original molecule: we identified heavy-atom sites bearing hydrogens, randomly selected a replaceable site, removed one hydrogen and appended a fragment from a curated list (\textbf{Supplementary Figure C.1}) by combining the parent and fragment molecules, forming a single bond at the attachment site. Each generated molecule is then sanitized using an open-source cheminformatics library (Rdkit) \citep{landrum2016rdkit} and discarded if this sanitization or valence checks fails. To favor synthetically accessible modifications – candidates with SCScore > 4.75 were discarded. Each molecule from the original dataset undergoes up to 20 such attempts (to compensate for rejected modifications) and up to 3 monovalent functionalization steps per attempt (number of attempts selected at random). The resulting molecules were canonicalized and filtered for duplicates. This ultimately yielded a final training corpus of 898,677 unique molecules. Given the scale of the dataset, although we do not have precise control over the quality of each generated molecule, the rigid SCScore cutoff and the sanitization and canonicalization protocols are expected to ensure that poor quality molecules are eliminated.

\begin{figure}[b!]
\centering
\includegraphics[width = 0.8\textwidth]{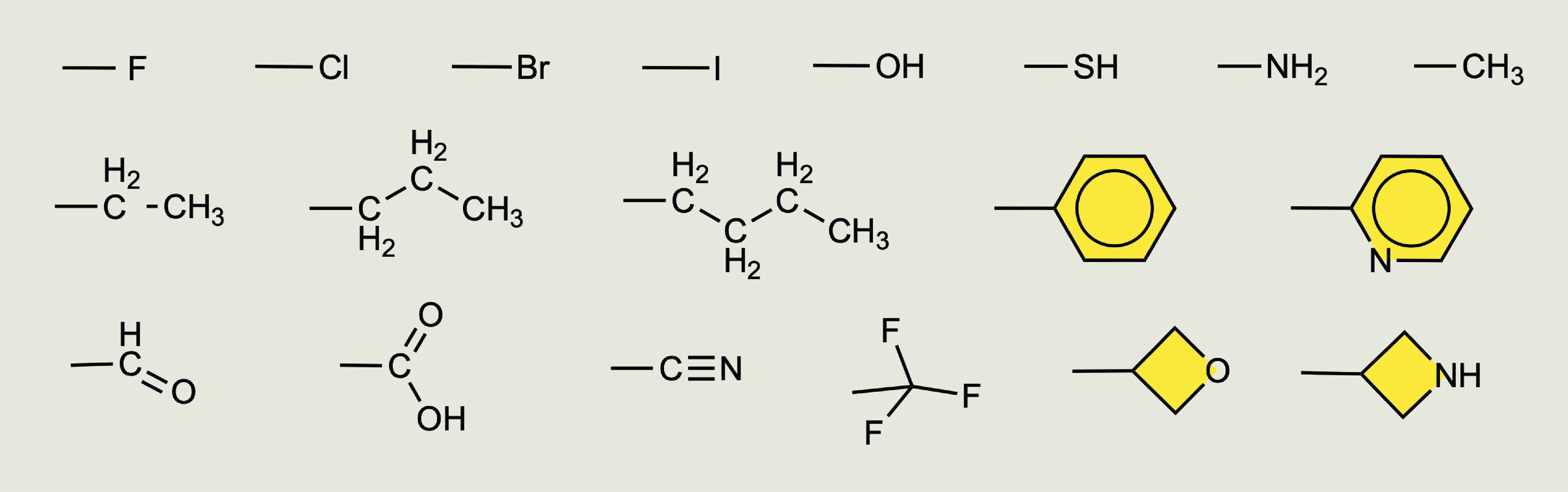}
\caption*{\textbf{Supplementary Figure C.1} List of fragments used to randomly functionalize linkers to expand the training corpus}
\end{figure}

\indent \indent \indent \textbf{Supplementary Figure C.2} shows distributions of molecular properties for the extended dataset post-functionalization (n = 200,000). The heavy atom count distribution is roughly unimodal, with most linkers in the medium-large regime (i.e., $\sim$ 20 – 30 heavy atoms) and relatively few short linkers (< 10 heavy atoms). Accordingly, the model is expected to learn medium-large linkers best, with limited support for very short linkers. The logP follows a bell-shaped curve – spanning both polar and hydrophobic linkers, and centered in a moderately lipophilic regime. Seeing no massive spikes at extreme ends is encouraging as it indicates that the data is not biased towards extremely polar or lipophilic molecules and covers a healthy range. The SCScore distribution is also reasonable – with the dataset explicitly biased away from extremely difficult linkers (i.e. scores > 4.75). There is a broad peak around 2 – 4, with very few molecules in the hyper-trivial regions (scores near 1). There is thus, a meaningful spread of synthetic difficulty – capturing what one would expect for MOF linkers. For rotatable bond counts (as a proxy for flexibility), the distribution is concentrated towards $\sim$ 5 - 10 rotatable bonds, with a tail > 20 and some entries close to 0. There is a mix of rigid, semi-rigid and flexible linkers (here, it is expected that functionalization would have increased these counts) – but the distribution is not overly skewed. The molecular weight distribution is also unimodal – capturing small bidentate to complex multidentate linkers. Overall, while functionalization gently skews some properties, the resulting distributions are smooth, unimodal and span chemical sensible ranges, yielding broad coverage biased towards medium-large, moderately flexible, moderately lipophilic, yet synthetically tractable linkers.\newline  

\renewcommand{\thealgocf}{C\arabic{algocf}}
\begin{algorithm}[H]
\caption{Dataset generation}
\label{alg:functionalization}
\KwIn{
    Parent molecule set $\mathcal{M}_0 = \{m_1, \dots, m_N\}$; \\
    Fragment set $\mathcal{F} = \{f_1, \dots, f_K\}$; \\
    Maximum attempts $A = 20$; \\
    Maximum substitutions per attempt $S_{\max} = 3$; \\
    Synthetic complexity (SCScore) threshold $\tau = 4.75$
}
\KwOut{Augmented molecule set $\mathcal{M}_{aug}$}
Initialize $\mathcal{M}_{aug} \leftarrow \emptyset$

\ForEach{$m \in \mathcal{M}_0$}{
    Identify eligible attachment sites:
    \[\mathcal{H}(m) = \{a \in \text{HeavyAtoms}(m) \mid \text{HydrogenCount}(a) \ge 1 \}
    \] 
    \For{$i = 1$ to $A$}{  
        $m' \leftarrow m$ \\
        Sample number of substitutions:
        \[
        s \sim \text{Uniform}\{1, \dots, S_{\max}\}
        \]
        \For{$j = 1$ to $s$}{  
            Randomly select attachment site:
            \[
            a \sim \text{Uniform}(\mathcal{H}(m'))
            \]          
            Randomly select fragment:
            \[
            f \sim \text{Uniform}(\mathcal{F})
            \]
            
            Remove one hydrogen from $a$
            
            Form single bond between $a$ and attachment atom of $f$
            
            Sanitize $m'$ (valence check, RDKit sanitization)
            
            \If{sanitization fails}{
                \textbf{break}
            }
        }
        \If{sanitization succeeds and SCScore$(m') \le \tau$}{
            Canonicalize $m'$;
            Add $m'$ to $\mathcal{M}_{aug}$}
    }
}
Remove duplicates from $\mathcal{M}_{aug}$
\Return $\mathcal{M}_{aug}$ 
\end{algorithm}

\begin{figure}[h!]
\centering
\includegraphics[width = \textwidth]{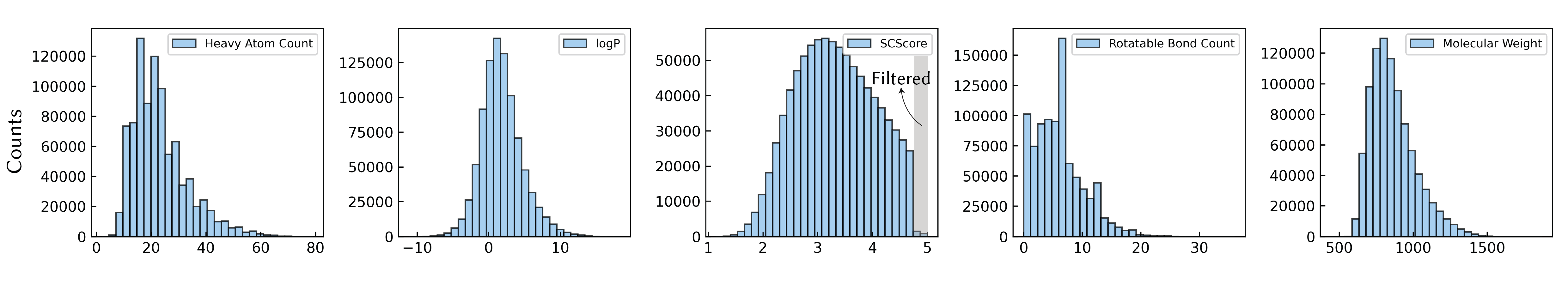}
\caption*{\textbf{Supplementary Figure C.2} Distributions of molecular properties – (from left to right) heavy atom count, logP, SCScore, rotatable bond count and molecular weight – for the extended dataset (n = 200,000 sampled randomly from a total 898,677 molecules). Note: The steep cliff observed in the SCScore distribution at large values is because we have filtered values exceeding 4.75 during the functionalization process.}
\end{figure} 

\indent \indent \indent \textbf{Supplementary Figure C.3} shows UMAP embeddings of linkers from the training dataset (Morgan fingerprints, radius 2, 2048 bits, n = 50000) coloured by logP, molecular weight and rotatable bond count – providing a complementary view of the corpus. The linkers occupy a single, coherent manifold with only a few small satellite regions, suggesting that the training data forms a connected chemical space rather than several isolated clusters. This suggests that the training distribution is continuous rather than being disjoint into isolated clusters (although some satellites do exist on the UMAP plot). Such distributions are favourable for models that rely on interpolation in the latent space. Variations in logP, size and flexibility are distributed across this manifold. There is some degree of qualitative visual coincidence between higher molecular weights and increased rotatable-bond counts, as expected for functionalised multidentate linkers. Regions with extreme values of these properties are present but relatively sparse and remain embedded within the main manifold, in agreement with the one-dimensional descriptor distributions (\textbf{Supplementary Figure C.2}).

\begin{figure}[h!]
\centering
\includegraphics[width = \textwidth]{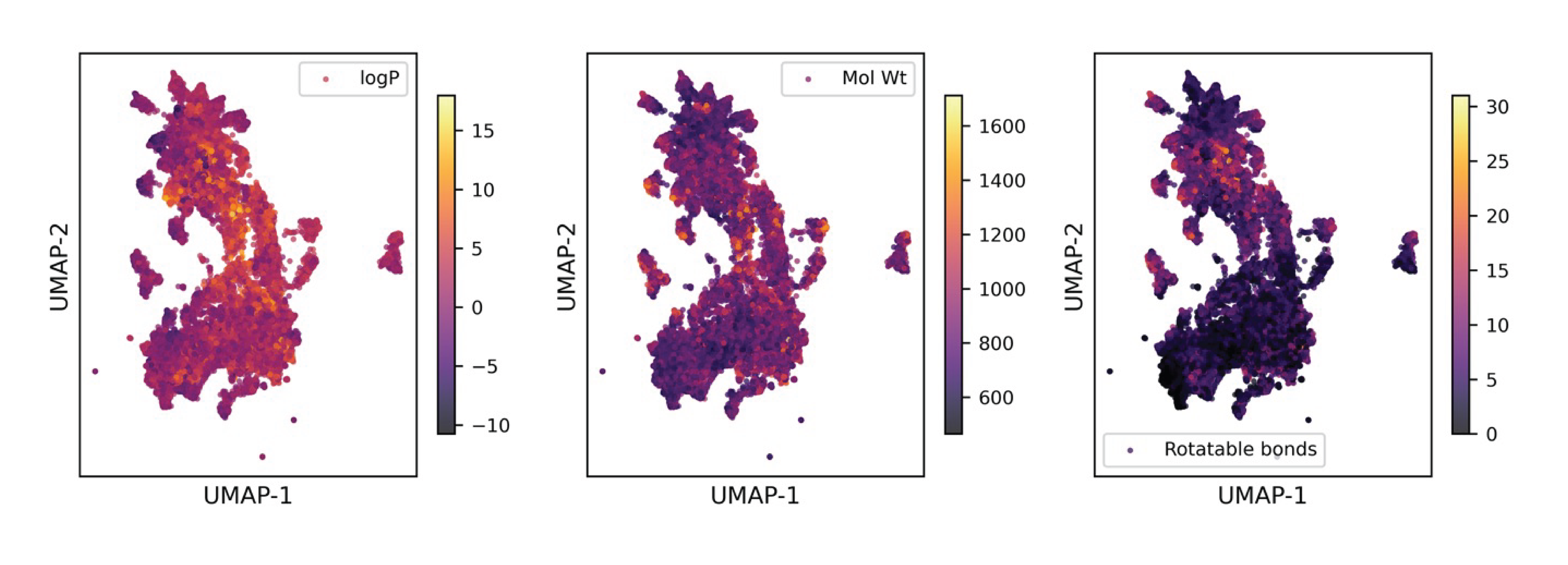}
\caption*{\textbf{Supplementary Figure C.3} UMAP visualization of the training corpus (n = 50,000, Morgan fingerprints, radius 2, 2048 bits). Colored by (left-right) logP, molecular weight and rotatable bond counts. Note: The colors have been normalized for each property class.}
\end{figure} 

\begin{figure}[h!]
\centering
\includegraphics[width = \textwidth]{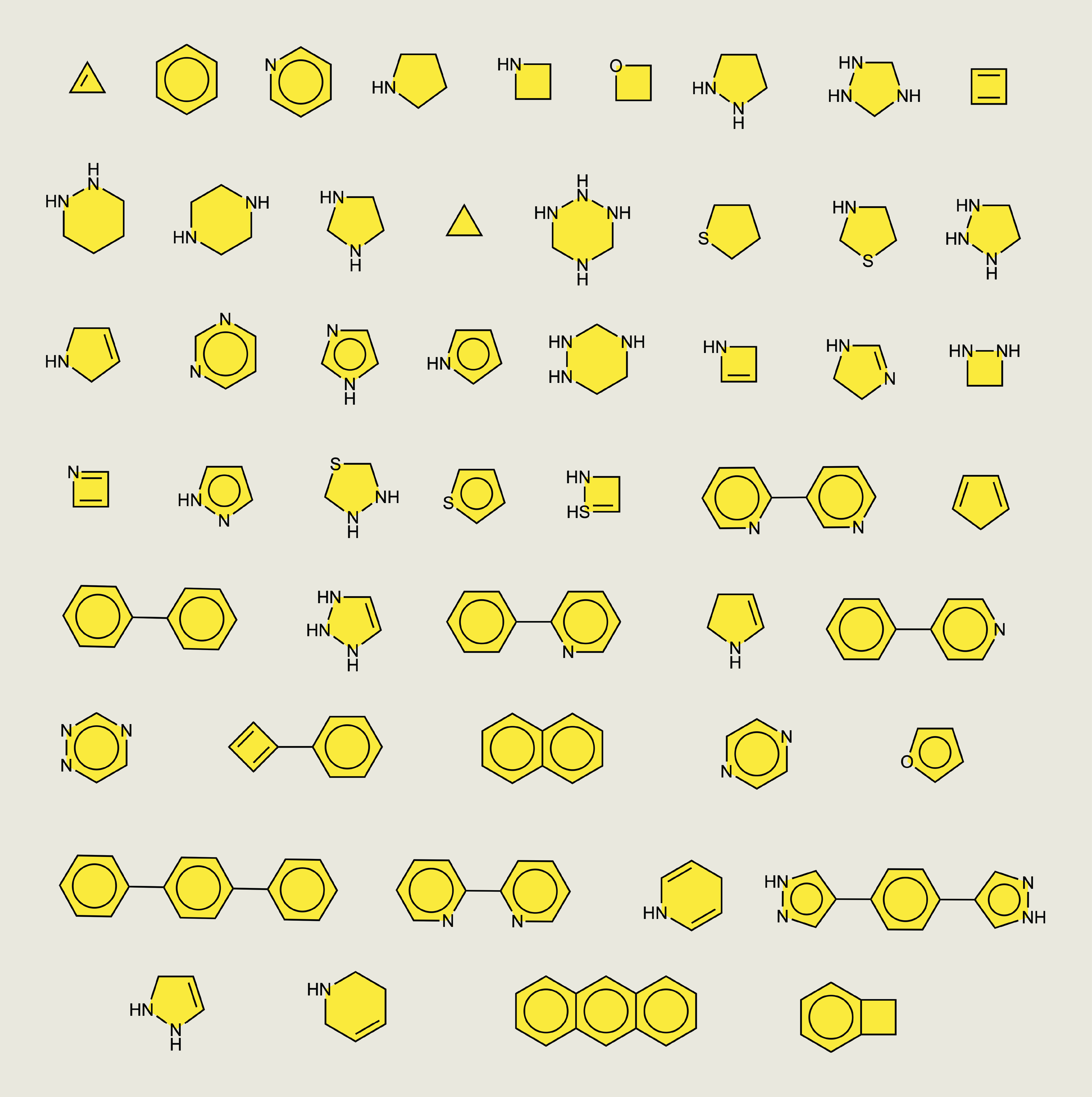}
\caption*{\textbf{Supplementary Figure C.4} The 50 most commonly observed Murcko scaffolds from the training dataset accounting for $\sim$ 30\% of the entire corpus. Ranked in decreasing order of prevalence from left-right, top-bottom.}
\end{figure} 

\indent \indent \indent Across the $\sim$ 900,000 linker corpus, we identified $\sim$ 103,000 unique Bemis-Murcko scaffolds – representing an $\sim$ 11\% uniqueness. Among these, the top 50 scaffolds (\textbf{Supplementary Figure C.4}) account for 30\% of all linkers, implying that $\sim$ 0.05\% of scaffolds – representing a small number of chemically simple cores – are reused, while the remaining cores have a longer tail of rarer, more complex motifs, not captured among the common cores. The commonly observed scaffolds are dominated by benzenoid and N-heterocyclic cores together with a few biaryl and fused aromatic systems, typical in MOF chemistry. The scaffold distribution provides frequent analogues on well-known MOF cores but also implies a bias towards these cores, with less statistical support for under-represented (complex) scaffold families. However, we anticipate that the scaffold-constrained design approach (discussed in the main manuscript) should be able to mitigate some of these biases.

%% file: sections/D.tex
Nexerra$^{R1}$ was trained over 172 epochs, with early stopping selecting a checkpoint around 120 epochs. \textbf{Supplementary Figure D.1} shows the training logs – which exhibit behavior consistent with stable training. The total and reconstruction loss rapidly decreases over the first few epochs. The slight increase in total loss in the 50 – 100 epoch range coincides with an increase in the $\beta$-weight assigned to the KL term in the loss function. Around the similar epoch-range, the KL loss exhibits a steep jump (corresponding to KL-free learning schedule) before gradually decreasing and stabilizing. The KL loss stabilizing at $\sim$ 25 is encouraging as it indicates that the model avoids posterior collapse, thanks to the $\beta$-warmup scheduler. This indicates that the latent space has meaningfully regularized and has not degenerated into a purely deterministic autoencoder. Parallelly, the per-token reconstruction accuracy jumps from $\sim$ 20\% to $>$ 90\% over the initial stage of training, following which it gradually stabilizes to $\sim$ 96.5\% over the training duration. For samples drawn during the training process, the internal diversity also stabilizes to slightly $>$ 0.925, which is close to what we observed for the training dataset, another encouraging sign. As for the novelty of these sampled molecules, it lies in the healthy 96 – 99.5\% range. The small fraction of exact reconstructions (causing the slight dip in novelty) is consistent with the models ability to reproduce training-like linkers while still generating predominantly novel molecules. Overall, the training logs support a non-degenerate representation of the MOF linker space that is capable of generating novel linkers. 

\begin{figure}[h!]
\centering
\includegraphics[width = \textwidth]{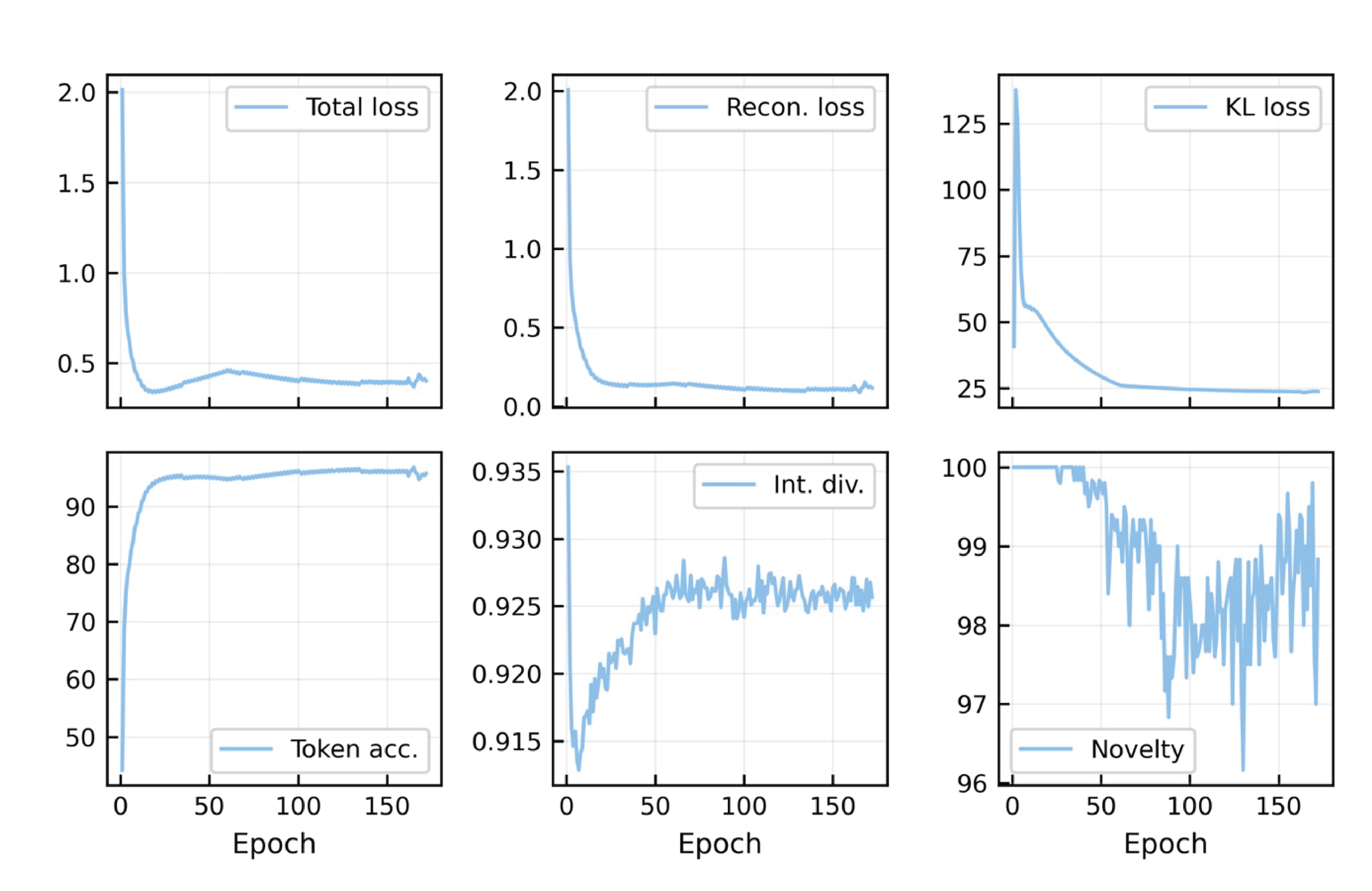}
\caption*{\textbf{Supplementary Figure D.1}. The training logs measured over 172 epochs indicate stable training. (First row, left to right): Total loss, reconstruction loss and KL loss. (Second row: left to right): Per-token reconstruction accuracy, internal diversity of sampled molecules and novelty of sampled molecules (n = 5000).}
\end{figure} 

\indent \indent \indent Beyond training metrics, a key test for a generative model is its ability to reproduce the property distributions of the data it is trained on. Using 10,000 molecules sampled from the model and 10,000 molecules randomly sampled from a previously unseen test set, \textbf{Supplementary Figure D.2} shows distributions for logP, fraction of sp$^3$ Carbons (Fraction Csp$^3$) and the total polar surface area (TPSA). For logP, the generated and test distributions are both unimodal and centred in the same region. The generated molecules show a slight right-shift towards more hydrophobic molecules with a longer tail – but the bulk of the probability mass overlaps (KL $\sim$ 0.055). For fraction Csp$^3$ – both distributions span the full 0-1 range. While the test set is more concentrated at the extremes, the generated distribution is flatter. That being said, the KL divergence of 0.078 is modest. As for the TPSA, both distributions have an excellent match with a very low KL divergence (KL $\sim$ 0.028). Taken together, the model is able to reproduce marginal distributions of key descriptors very well. There are minor deviations, but these effects are confined to distribution tails. 

\begin{figure}[h!]
\centering
\includegraphics[width = \textwidth]{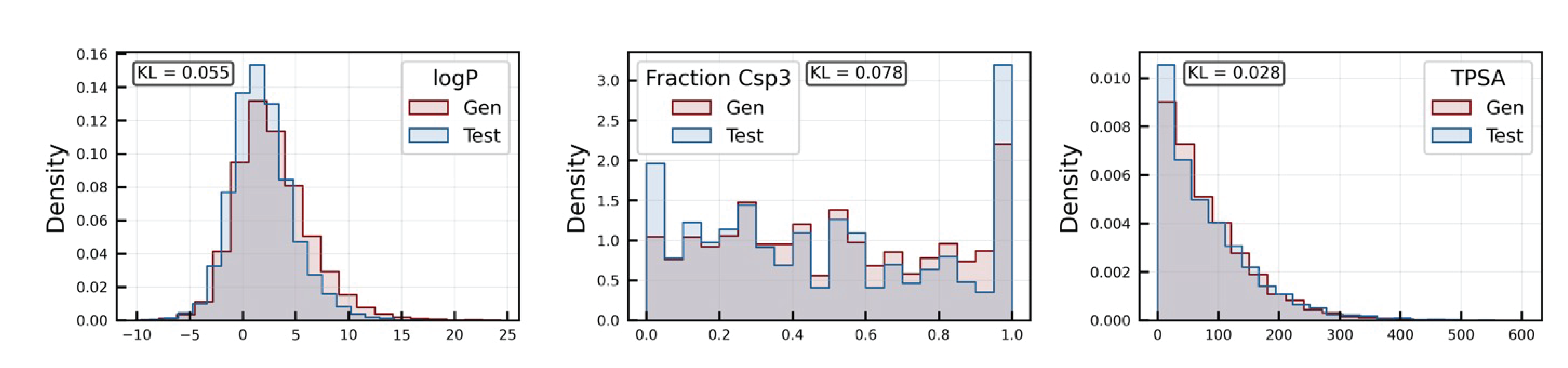}
\caption*{\textbf{Supplementary Figure D.2}. Distributions for (left to right) logP, fraction Csp$^3$ and TPSA for generated molecules (n = 10,000) and a non-redundant subset of the test set (n = 10,000). The annotated text represents the KL divergence between the two distributions. A KL divergence of 0.0 would indicate a perfect match (the ideal case).}
\end{figure} 

\indent \indent \indent Next, to assess the continuity of the learned latent representation, we projected 50,000 training embeddings and 49,950 samples drawn from the latent prior into a shared UMAP embedding (\textbf{Supplementary Figure D.3}). The generated samples closely overlap the empirical manifold without forming explicitly disjoint clusters or major regions outside the training manifold. This suggests that the VAE prior captures a coherent chemical distribution and that unconstrained sampling remains within the support of the learned latent space. 

\begin{figure}[h!]
\centering
\includegraphics[width = 0.6\textwidth]{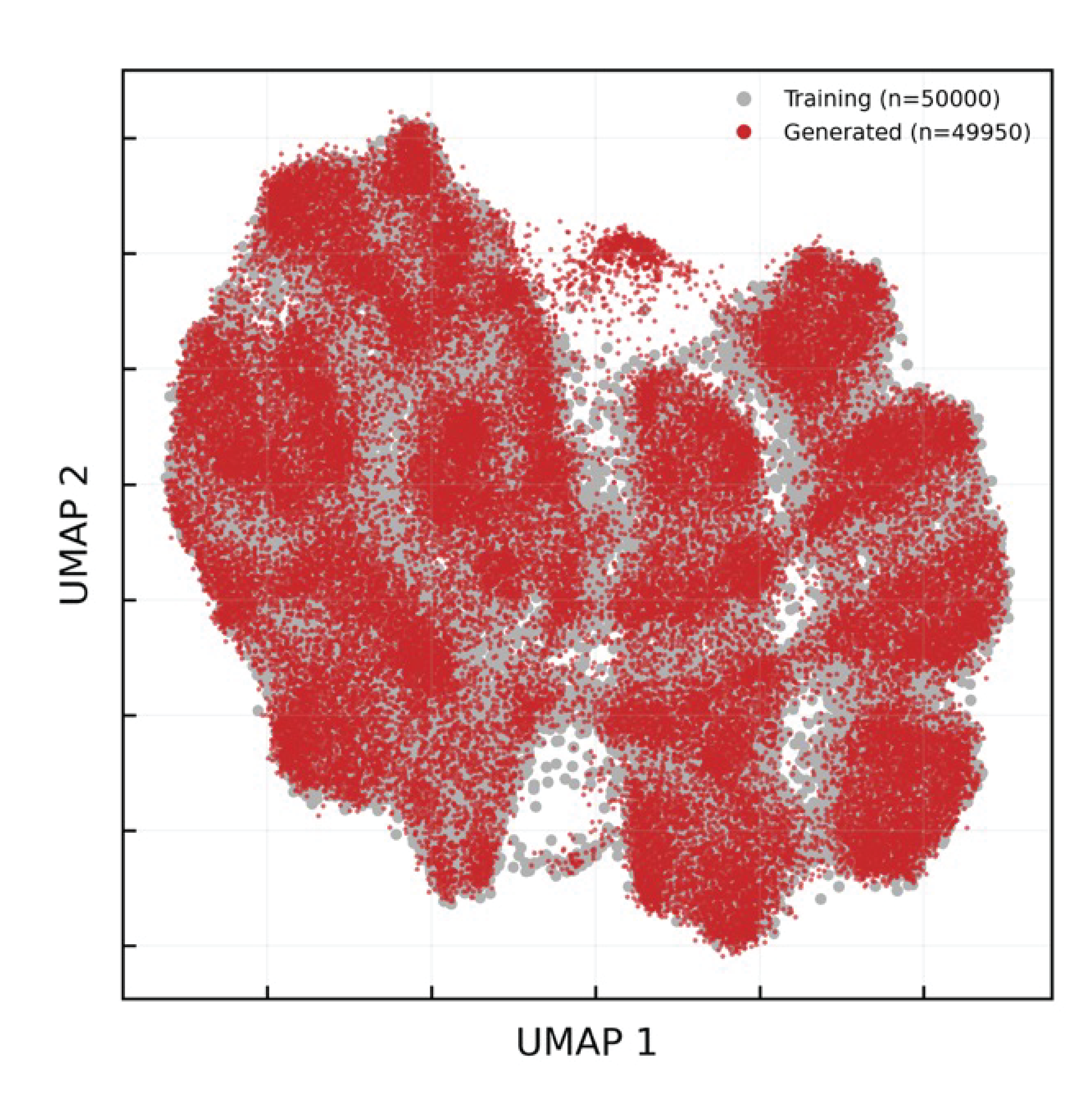}
\caption*{\textbf{Supplementary Figure D.3}. UMAP visualization of the latent embeddings of training molecules (gray, n = 50,000) and generated molecules (red, n = 49,950). The generated samples closely overlap the training manifold, although, there is mild clustering of generated molecules in some pockets. Additionally, there are no major pockets outside the training manifold, suggesting that sampling remains within the learned latent space.}
\end{figure} 

\indent \indent \indent \textbf{Supplementary Figure D.4} shows examples that illustrate the structures Nexerra$^{R1}$ can access under free sampling from the prior. The model readily proposes multi-functional, high-connectivity linkers with extended aromatic cores and dense substitution patterns. This complexity reflects both the expressivity of the language model and its large context window (up to 102 tokens), which allows it to coordinate long-range patterns such as symmetry, repeating units and balanced placement of coordination groups along the backbone. However, we must stress that these are representative high-quality samples selected from a much larger pool: in unconstrained free sampling, the same expressive capacity also produces many molecules that are synthetically unappealing, poorly suited to MOF linkers, or simply off-target for the design objective. In other words, the model’s raw generative power does not automatically translate into an efficient ‘hit rate’ for good candidates. In practice, we therefore rely on seed-guided design, where a chemically reasonable linker template or scaffold is provided and the model is used to explore the chemical vicinity of this scaffold in the latent space pursuant to a reward function, to focus generation onto promising regions of linker space and substantially improve the efficiency of discovering useful linkers. 

\begin{figure}[h!]
\centering
\includegraphics[width = \textwidth]{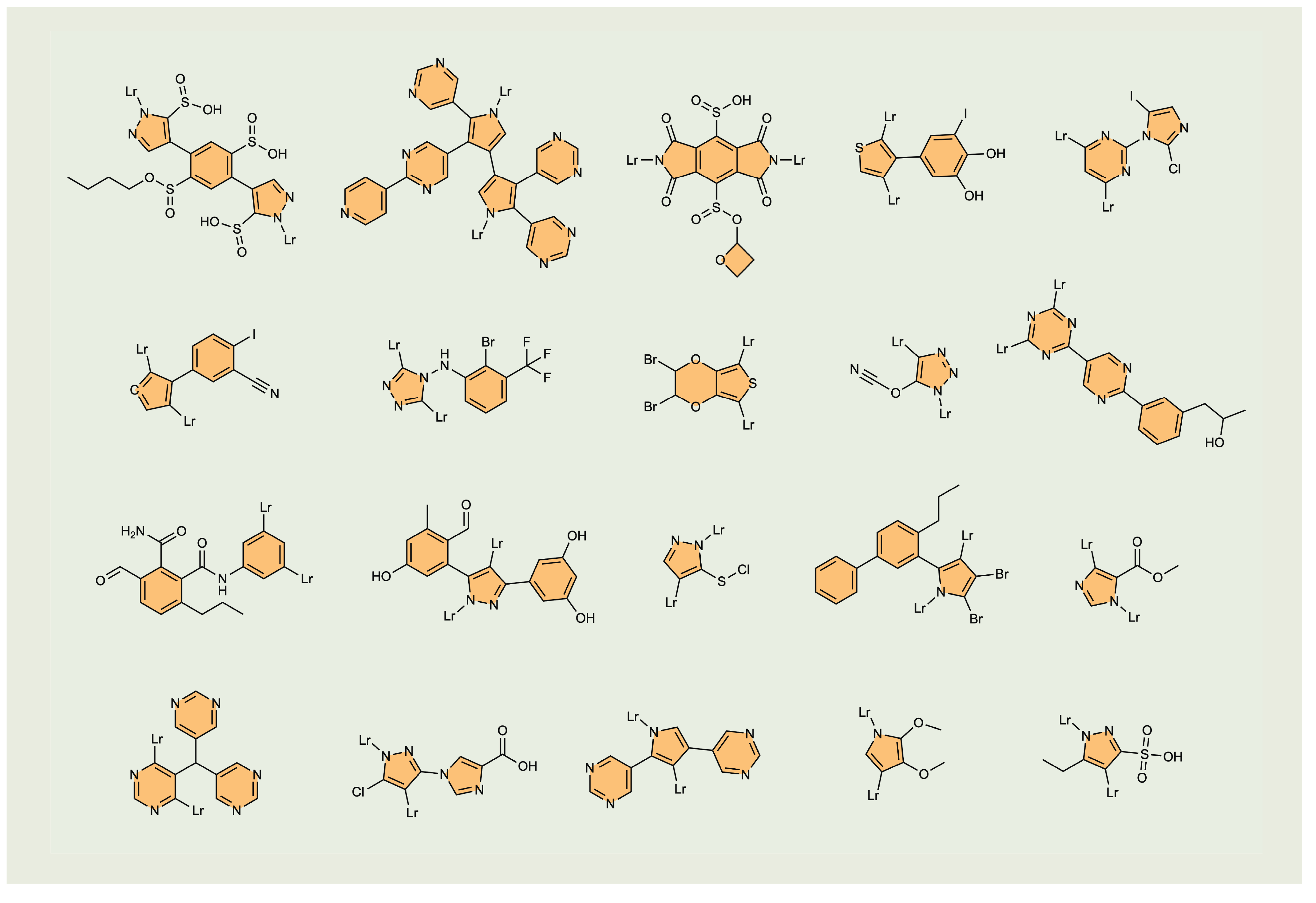}
\caption*{\textbf{Supplementary Figure D.4}.
Illustrative examples of molecules sampled freely from the prior of Nexerra$^{R1}$. The model is able to generate multi-functional, complex molecules with polyaromatic cores and dense substitution patterns. However, owing to the high expressivity of the CLM, under unconstrained sampling, the ‘hit rate’ for desirable linkers is low. Thus, seed-guided design serves as a more optimal route towards the discovery of exceptional linkers.}
\end{figure} 

\indent \indent \indent Finally, using the 10,000 linkers sampled from the latent prior, we computed standard distributional diagnostics. All generated molecules were chemically valid (validity = 1.00; as expected when using SELFIES), with negligible duplication (uniqueness = 0.99), indicating that the decoding is stable, with no mode collapse. Internal diversity remained high (IntDiv = 0.925), comparable to that of the training distribution, suggesting broad exploration of linker chemotypes. The average similarity to the nearest  molecule (SNN) = 0.26, reflecting moderate structural proximity without memorization; this is also confirmed by the latent space embeddings (\textbf{Supplementary Figure D.3}) which show that the sampled molecules occupy the same chemical space as the training dataset. The scaffold diversity (Scaff = 0.92) confirms that linker generation is not confined to a narrow subset. Put together, these metrics indicate that the learned latent prior captures a coherent and chemically diverse distribution, providing a stable foundation for subsequent distributional targeting via flow.

%% file: sections/E.tex
Once generated, molecules are subjected to standard sanitization checks using RDKit \citep{landrum2016rdkit}; these include valency checks, aromaticity and ring assignment checks. Should a generated molecule fail this, it is immediately discarded from any further consideration. This is however, rare, considering we use SELFIES representation for the model which guarantees validity.\newline   
\indent \indent \indent To coordinate with the metal-containing cluster (i.e., the node), the linker should have coordinating moieties. In practice, these are usually groups such as carboxylates, phosphonates, sulfonates or azolates. However, due to the nature of the MOF assembly algorithm \citep{colon2017topologically} that we use, we instead denote these moieties using an ‘[Lr]’ token. Now, in addition to having these ‘[Lr]’ tokens, they must be placed onto the generated linker is a manner that ensures geometric and topological compatibility with the target net.\newline 
\indent \indent \indent First, we remove all existing ‘[Lr]’ tokens; in case this token is on a ring system, it is substituted with a Nitrogen atom. This resulting scaffold is then geometry optimized. The number of coordination handles to be placed either depend on the input seed (i.e., to match the coordination of the seed) or alternatively, can be specified by the user. In principle, the algorithm is agnostic to the number of such handles. Next, we embed the linker in Euclidean space and define candidate attachment sites as the set of carbon atoms with at least one removable hydrogen atom. Let $S =\{s_1,...,s_n \}$ denote a set of 3D coordinates of these carbon atoms. In order to place $k$ anchors, we enumerate all possible $k$ candidate sites. For each candidate set $S_i=\{s_{i_1},...,s_{i_k}\}$, we calculate the mean pairwise separation as,
\begin{equation}
    \Phi = \frac{2}{k(k -1)}\sum_{a < b}\|r_{i_a} - r_{i_b}\|
    \tag{E.1}
\end{equation}

\indent \indent \indent We select the configuration that maximizes $\Phi$. This favors spatially well-separated anchors with good linker geometries and low steric crowding. This works especially well for ditopic and tritopic linkers. Following the selection of these candidate sites, we remove one hydrogen from these specified Carbons and replace it with an ‘[Lr]’ token. \textbf{Supplementary Figure E.1} describes the linker optimization workflow adopted. 

\begin{figure}[h!]
\centering
\includegraphics[width = \textwidth]{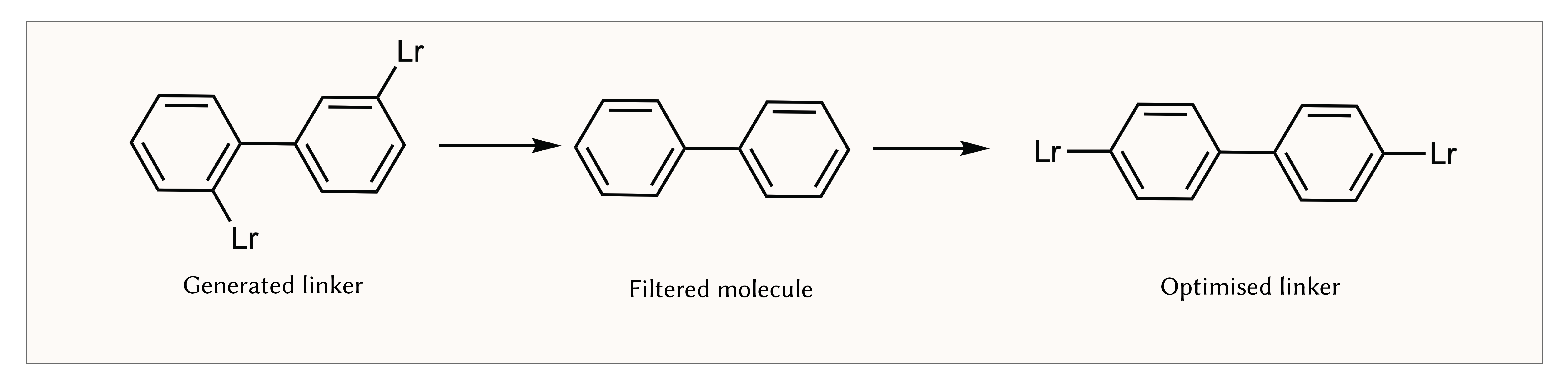}
\caption*{\textbf{Supplementary Figure E.1}.
Starting with a generated linker, it is stripped out its ‘[Lr]’ handles, and optimised. Next, we collect a set of candidate attachment sites (described above) and select those sites which maximise the spatial separation between anchor points. We append the required ‘[Lr]’ handles to these selected sites. This workflow is in principle, agnostic to the number of handles.}
\end{figure} 

%% file: sections/F.tex
The generative capabilities of Nexerra$^{R1}$ do not extend to the efficient discrimination of useful linkers – i.e., linkers with favorable properties – against linkers that are sub-optimal. Even with the addition of the flow model (\textbf{Supplementary Information Section I}), we are at best able to favor linker sampling against certain target properties (for instance, linker length). For most applications, we need to optimize several, often competing properties. We thus rely on reward functions that score linkers based on their utility for the application of interest. In this context, reward functions offer several advantages: (i) they are modular, i.e., they can be easily formulated for an application of interest to screen for desirable linkers; (ii) they are computationally cheap and do not require any additional model training; and (iii) they can span arbitrary complexity. On the downside, these functions rely on several simplifications and assumptions, and on the whole are tough to generalize. Functions formulated for a system of interest, may not necessarily work well for other systems. Nonetheless, the flexibility on offer enables access to vast design spaces that in principle, span multiple application landscapes.\newline
\indent \indent \indent Here, as case studies, we formulate two reward functions targeting gas storage and biocompatibility. Before proceeding to the formulation of these functions, we preface the discussion by clarifying that these functions operate completely at the linker-level and do not explicitly account for metal-node and topology effects, host-guest interactions, and framework flexibility. These characteristics undoubtedly play critical roles in the overall performance, and can be evaluated using geometric analysis and molecular simulations on assembled frameworks. The goal here is to develop computationally lightweight structural filters to bias generation towards desirable chemotypes.\newline 

\noindent \textbf{F.1. Gas storage.} In the design of adsorbents for gas storage (e.g., H$_2$, CH$_4$), the deliverable capacity is a critical design parameter \citep{chen2020balancing}. Here, the gravimetric uptake – mass of adsorbate taken up per unit mass of adsorbent – and the volumetric uptake – mass of adsorbate taken up per unit volume of adsorbent – are two equally important, yet competing metrics \citep{chen2020balancing, chen2025topology}. Design strategies need to focus on concurrently optimizing both objectives. At the linker level, several structural characteristics influence framework-level gas storage capabilities: (i) anchor-to-anchor span, which modulates the node-node separation and the pore size; (ii) molecular weight, which controls framework density; (iii) rigidity, which impacts disorder and framework collapse; and (iv) linker symmetry, which impacts the packing efficiency.\newline
\indent \indent \indent Starting with gravimetric storage, the capacity benefits by maximizing accessible void volume per unit mass. To approximate this at the linker-level, we define a mass-normalized span descriptor as:
\begin{equation}
    d_{mass-norm-span} = \frac{d_{anchor - anchor}}{MW}
    \tag{F.1}
\end{equation}

\noindent where, $d_{anchor-anchor}$ is the maximum anchor – anchor distance and MW is the molecular weight of the linker (excluding anchors). $d_{mass-norm-span}$ serves as a proxy for structural expansion per unit mass.  This proxy – which is intended as a purely relative descriptor – biases generation towards longer, lightweight linkers. At the same time, such linkers tend to exhibit a high-degree of flexibility, which makes ultraporous frameworks susceptible to potential pore collapse. We therefore introduce a second descriptor $d_{rigidity}$ that penalizes flexible linkers that may promote framework collapse. The rigidity proxy is defined as follows:
\begin{equation}
    d_{rigidity} = 0.5 \times RBC + 0.5 \times N_{C_{sp^3}}
    \tag{F.2}
\end{equation}

\noindent where, $RBC$ is the fraction of the rotatable bonds in the linker and $N_{C_{sp^3}}$ is the fraction of $sp^3$ Carbons in the linker. Put together, the gravimetric reward ($R_{grav}$) is defined as follows:

\begin{equation}
 R_{grav} = \alpha _1 D(d_{anchor-anchor}) + \alpha _2 D(d_{rigidity}) + \alpha _3 D(d_{mass-norm-span})
 \tag{F.3}
\end{equation}

\noindent where, $D$ denotes bounded desirability functions that prevent extreme values from distorting the reward scores while enabling smoother interpolations; $\alpha_1, \alpha_2, \alpha_3$ denote the relative weights assigned to each term. In the case of the mass-normalized span descriptor, $D$ is an increasing sigmoidal function centered around the $75^{th}$ percentile of the distribution from the sampled training set (\textbf{Supplementary Figure F.1.a}). Due to the strong regularization effect of this descriptor that penalizes heavy linkers, it is assigned a lower weightage of 0.2. For the flexibility descriptor, $D$ is a decreasing sigmoidal function centered around 0.5, with a width of 0.2. For the span descriptor, $D$ is an increasing sigmoidal function centered around 16 Å, with a width of 4 Å. Overall, $R_{grav}$ encodes structural priors that support lightweight, long-span, shape-persistent linkers. \textbf{Supplementary Figure F.1.b} shows the $R_{grav}$ distribution of linkers from the training set (n = 25,000). The distribution has a median reward of 0.52 (p75 = 0.58, p25 = 0.43). Crucially, the distribution is not skewed to either extreme confirming that the rewards are not saturated; instead, the rewards span a healthy range of (0.2, 0.8). The absence of sharp cutoffs or collapsed modes indicates that the reward landscape is well-conditioned and preserves meaningful variance across chemically diverse linkers. The function should ensure that structural modifications produced graded changes in the reward – consistent with flow-based optimization.

\begin{figure}[h!]
\centering
\includegraphics[width = \textwidth]{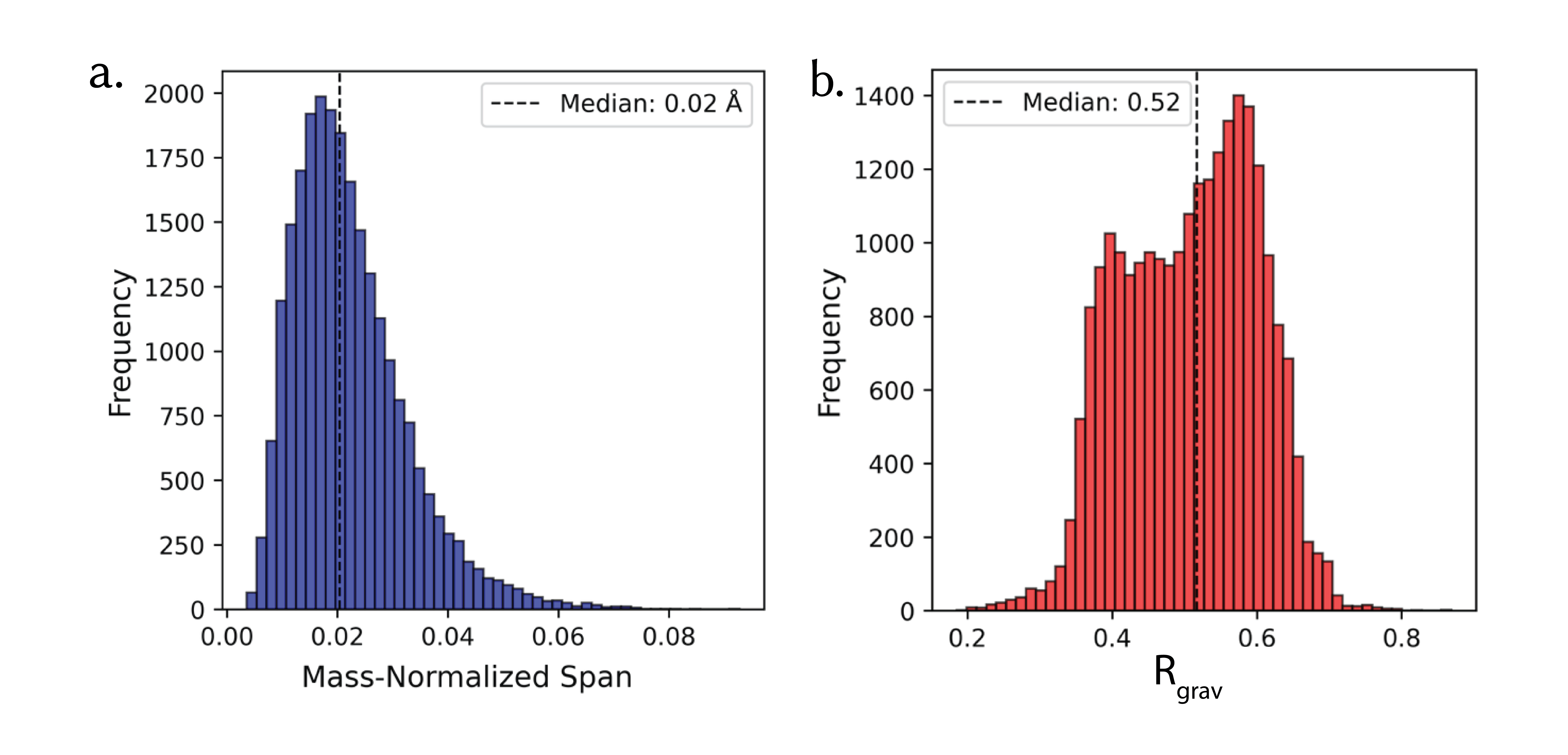}
\caption*{\textbf{Supplementary Figure F.1 a}.Distribution of the mass-normalized span for molecules from the training set (n = 25,000). Median = 0.02 Å mol g$^{-1}$; p75 = 0.03 Å mol g$^{-1}$ and p25 = 0.01 Å mol g$^{-1}$. The increasing sigmoidal desirability function is centred around p75, with a width of 0.5 $\times$ (p75 – p25) to favour linkers with favourable mass-normalized spans. \textbf{b.} Distribution of the gravimetric reward score ($R_{grav}$) for molecules from the training set (n = 25,000). Median = 0.52; p75 = 0.58 and p25 = 0.43. Notably, the distribution is not skewed towards either extreme bound – which would have been indicative of collapse.  The distribution spans almost the entire range and we also observe a smooth gradient towards higher scores, suggesting that a non-trivial fraction of the training linkers achieve high scores [$\alpha _1= \alpha _2= 0.4; \alpha _3 = 0.2]$.}
\end{figure} 

\indent \indent \indent \textbf{Supplementary Figure F.2} shows the $R_{grav}$ scores computed for linkers corresponding to some of the top-performing MOFs for gravimetric CH$_4$ storage. The scores assigned to each linker roughly correspond to the observed CH$_4$ storage trends. Notably, the linker of CU-6 – a MOF recently reported by some of us \citep{chen2025topology} – which has the highest gravimetric CH$_4$ storage capacity to date shows the highest relative $R_{grav}$ score. We however stress that there are several aspects relating to metal-node effects, topology and volumetric packing that are not captured here; this remains a computationally-efficient proxy for discriminating linkers.

\begin{figure}[h!]
\centering
\includegraphics[width = \textwidth]{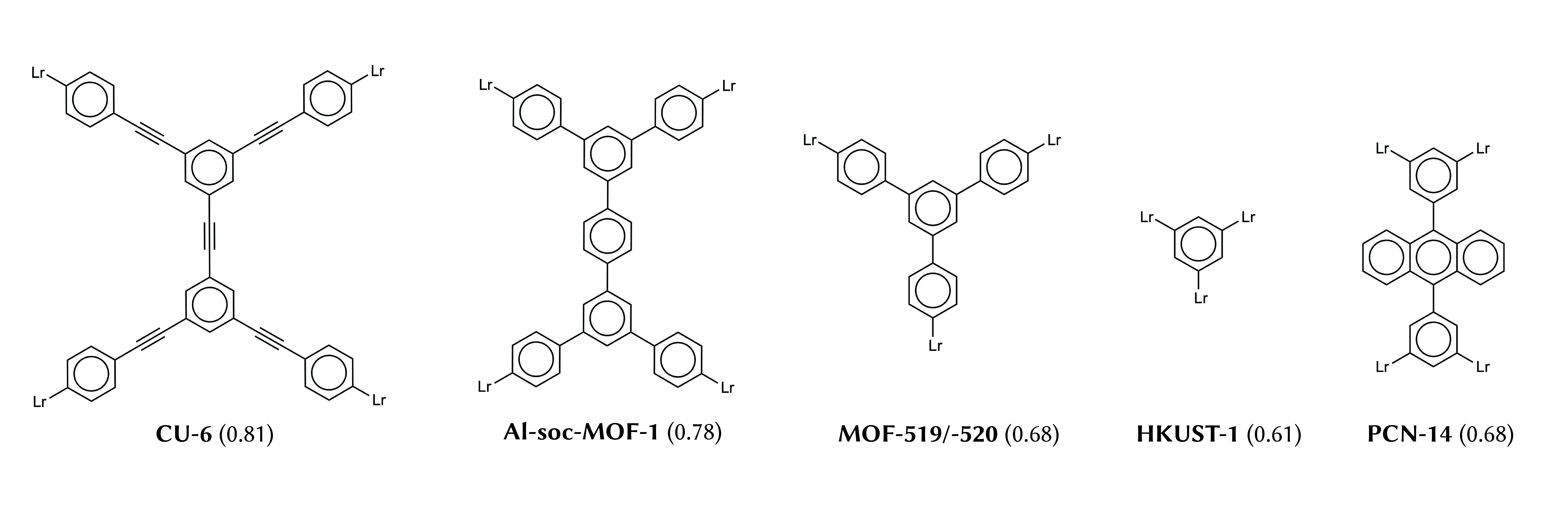}
\caption*{\textbf{Supplementary Figure F.2.} $R_{grav}$ scores for linkers belong to top-performing MOFs for gravimetric CH$_4$ storage. The ranking of benchmark linkers is broadly consistent with reported trends. However, for a complete assessment, it is crucial to account for metal-node effects, topology and volumetric packing. As an illustrative example, while MOF-519 and MOF-520 share the same linker, they differ in coordination environments that in-turn have a significant impact on storage capacities.}
\end{figure} 

\indent \indent \indent Moving to volumetric storage, the goal is to balance porosity with the framework density. Excessively long, lightweight linkers which favor gravimetric storage, will lower the volumetric capacity due to reduced packing density. Accordingly, a balanced reward function needs to account for: (i) anchor – anchor span; (ii) rigidity; (iii) symmetry of the linker (or a proxy-thereof); and (iv) intermediate molecular weights that avoid excessively heavy (reduces gravimetric capacity) and light (reduces volumetric capacity) linkers. The balanced gas storage reward ($R_{gas}$) is defined as follows:

\begin{equation}
    R_{gas} = \beta _1 D(d_{anchor - anchor)} + \beta _2 D(d_{rigidity}) + \beta _3 D(d_{symmetry}) + \beta _4 D(d_{MW}) 
    \tag{F.4}
\end{equation}

\noindent where $d_{symmetry}$ is a measure of similarity of the chemical environment around each anchor using Morgan fingerprints (a proxy for linker symmetry) and $d_{MW}$ is the molecular weight of the linker passed through a Gaussian desirability function that penalizes extremely heavy and lightweight linkers while rewarding linkers with intermediate molecular weights (\textbf{Supplementary Figure F.3.a}). Each component is bounded between [0.05,0.95] and the parameters $\beta _1 - \beta _4$ are tunable. $R_{gas}$ is similarly bounded between [0.05,0.95], with higher values suggesting an increased applicability for gas storage. We note here that several parameters (especially the Gaussian desirability function) may need to be tuned to the system of interest.\newline 
\indent \indent \indent The parameters $\beta _1 - \beta _4$ were empirically optimized against known linkers belonging to high-performing MOFs, with weights selected to preserve qualitative ranking of linkers corresponding to benchmark MOFs. At values of $\beta_1$ = 0.2; $\beta_2$ = 0.2; $\beta_3$ = 0.4 and $\beta_4$ = 0.2, the reward score trends were found to reasonably correlate to experimental data. \textbf{Supplementary Figure F.3.c} shows the distribution of $R_{gas}$ values using these parameters when evaluated on 25,000 molecules from the training dataset. The distribution is broad (with a median at 0.43) and similar to the $R_{grav}$ distribution, does not saturate at either extreme. There is a weak bimodal character, with a secondary shoulder at higher values, suggesting a partitioning of the linker space into two regimes. The larger class is geometrically sub-optimal (\textbf{Supplementary Figure F.3.b}), while the smaller class is primarily driven by symmetry and rigidity components of the score. This is by design, considering that reticular assembly of frameworks of interest favor linkers with symmetric anchor environments and lower conformational flexibility yet having a reasonable span. 

\begin{figure}[h!]
\centering
\includegraphics[width = \textwidth]{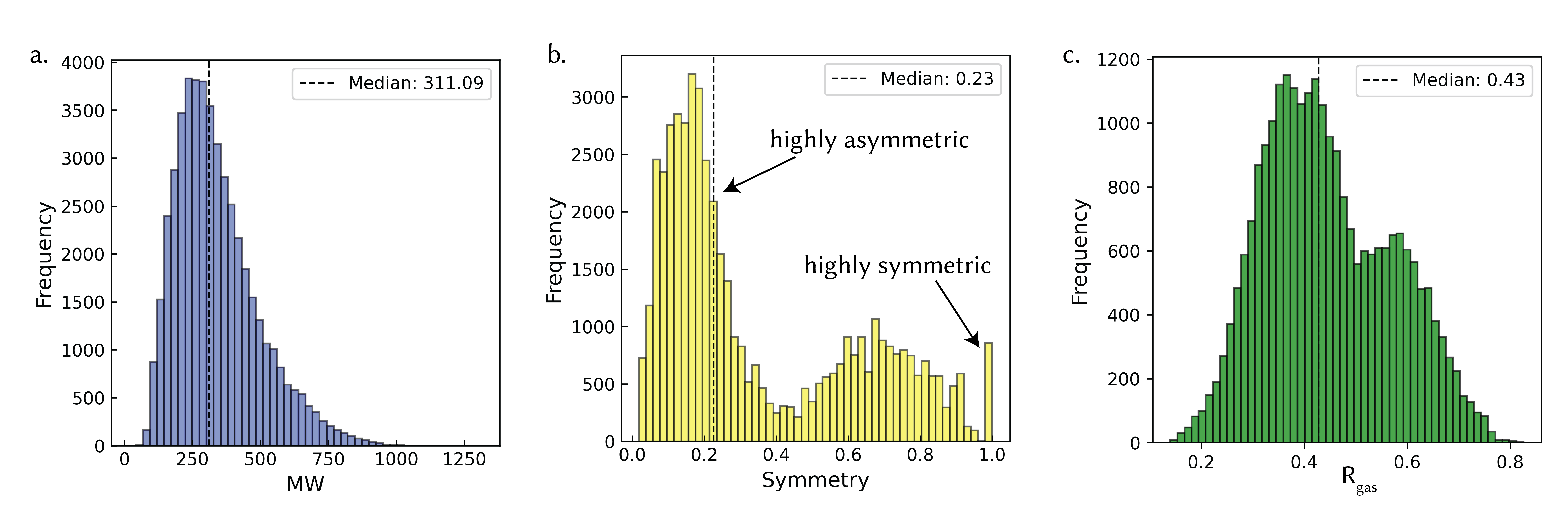}
\caption*{\textbf{Supplementary Figure F.3 a}. Distribution of the molecular weight of molecules from the training dataset (n = 50,000). The distribution follows an expected trend with a mode around 250 g mol$^{-1}$, median of 312 g mol$^{-1}$ and a long tail that extends past 1250 g mol$^{-1}$. Ideal linkers for gas storage should adopt intermediate molecular weights – i.e., not too light to favor low framework densities yet not too heavy to favor high densities. We thus define a Gaussian desirability function (centered around 320 g mol$^{-1}$) with a width of 140 g mol$^{-1}$ that rewards these intermediate values while penalizing extremes. \textbf{b.} Distribution of the $d_{symmetry}$ (n = 50,000) and \textbf{c.} $R_{gas}$ scores (n = 25,000) of molecules from the training dataset. $d_{symmetry}$ has a strong bimodal distribution suggesting the partitioning of the linker space into two classes: linkers with sub-optimal symmetry and linkers with optimal symmetry. Only 1.6\% of the linkers attain the maximum $d_{symmetry}$. For $R_{gas}$, the distribution is broad (median = 0.43) with a weak bimodal character. Neither extreme is saturated – confirming that there is no collapse. The weak bimodal character stems from the partitioning of the linker space described above. Here, the smaller class indicates more favorable symmetry and rigidity yet having a reasonable span and molecular weight.}
\end{figure} 

\indent \indent \indent \textbf{Supplementary Figure F.4} shows the $R_{gas}$ scores computed for linkers corresponding to some of the top-performing MOFs for gravimetric CH$_4$ storage. Again, the scores assigned to each linker correlate well with the observed trends. In this instance, the linker of Al-soc-MOF-1 \citep{alezi2015mof} shows a higher relative $R_{gas}$ score compared to the linker of CU-6. The linker in PCN-14 shows the highest relative score; while the MOF itself is not the best performing, its performance is competitive nonetheless. Yet again, we stress that there are several aspects not captured here; we remind the reader that this remains a computationally-efficient proxy for discriminating linkers.\newline 

\begin{figure}[h!]
\centering
\includegraphics[width = \textwidth]{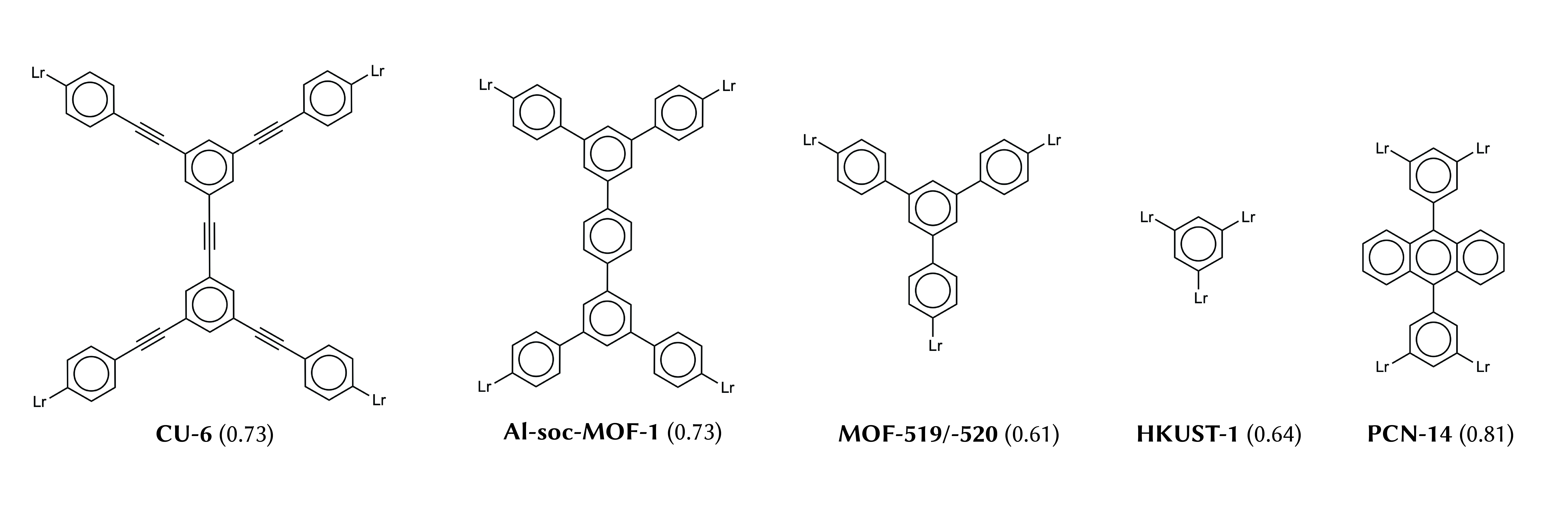}
\caption*{\textbf{Supplementary Figure F.4}. $R_{gas}$ scores for linkers belong to top-performing MOFs for CH$_4$ storage. There is a reasonable correlation between the score assigned to the linker and the reported gravimetric and volumetric storage capacities of the MOF. However, for a complete assessment, it is crucial to account for metal-node effects, topology and volumetric packing.}
\end{figure} 

\noindent \textbf{F.2. Biocompatibility.} For passive drug delivery applications, the goal is not only to identify synthetically plausible linkers, but to prioritize chemotypes that remain safe, soluble, and functionally effective after framework assembly. Consistent with the gas-storage analysis above, we preserve the geometric approach of linker-level screening while introducing safety, performance and compatibility controls that are directly relevant to biomedical deployment.\newline 
\indent \indent \indent The rationale of this reward function is again, modularity, focusing on: (i) safety terms penalizing metals and linker chemistries or fragments associated with high acute-toxicity risk; (ii) performance emphasising linker span, solubility and accessible porosity properties for loading space, which include pore-limiting diameter (PLD) and accessible surface area (ASA) or derived approximations if unavailable; (iii) benign-linker terms suppressing strongly interacting or reactive motifs that can perturb therapeutic cargo; (iv) stability prioritising candidates less likely to trigger burst release, premature linker failure, or charge-driven instability under physiological conditions; finally (v) penalty element penalising linkers with known toxicity warnings and extreme charge states.
We define a biocompatibility objective as follows:

\begin{equation}
    R_{bio}  =  \gamma_1 S + \gamma_2 P + \gamma _3 B + \gamma _4 T - P
    \tag{F.5}
\end{equation}

\noindent where $S$, $P$, $B$, and $T$ denote safety, performance, benign-linker behaviour, and stability sub-scores, respectively, and $P$ is an explicit penalty for warnings and extreme charge states. All scores and components range between $[0,1]$.\newline 
\indent \indent \indent The safety block (S) combines: (i) metal-node safety from curated LD50-based priors reported by \cite{menon2025guiding}, (ii) toxicity-space proximity derived from acute-toxicity chemistry, and (iii) formal-charge safety,

\begin{equation}
    S = s_1.M_{metal-toxicity} + s_2.T_{tox-space} + s_3.Q_{charge}
    \tag{F.6}
\end{equation}

The toxicity-space term was built from descriptor and fingerprint neighbourhoods obtained from median lethal dose (LD50) readouts for 35,300 compounds in TOXRIC Database \citep{wu2023toxric}. 55-feature standardized numeric manifold (RDKit descriptors) and ECFP4 fingerprint manifold (2048 bits) were blended by learned block weights: numeric 0.62, fingerprint 0.38. Toxicity severity was weighted across United Nations Globally Harmonized System of Classification and Labelling of Chemicals (UN GHS) classes (Class 1 as the most severe to Class 5 associated with least toxic compounds) so that proximity (k-NN) to highly toxic regions is penalized more strongly.\newline
\indent \indent \indent Toxicity-space was validated using several MOF linkers and other chemicals that were reported to have been experimentally tested, including PCN-222 (Tetrakis (4-carboxyphenyl) porphyrin (TCPP) linker), UiO-66 (terephthalic acid linker), NU-1000 (Pyrene tetrabenzoic acid) and ZIF-8 (2-imidazoline, featuring known LD50 oral rat toxicity at $\sim$ 1,400 mg kg$^{-1}$).\newline 

\noindent The performance block (P) combines linker length, solubility, and surface/loading proxies,
\begin{equation}
    P = p_1(d_{anchor-anchor}) + p_2(d_{solubility}) + p_3(d_{loading})
    \tag{F.7}
\end{equation}

\noindent where linker length $d_{anchor-anchor}$ is estimated from anchor-to-anchor separation and heavy-atom span; $d_{solubility}$ is encoded using LogS together with a bounded LogP window; $d_{loading}$ is a surface/loading proxy which uses accessible surface area (ASA) and pore limiting diameter (PLD) when available, with deterministic linker-level proxies otherwise. Pore accessibility and packing constraints remain fundamental for biocompatibility applications as well.\newline 
\indent \indent \indent The benign-linker block (B) combines formal-charge moderation, a payload-interaction window (LogP and TPSA), and structural alert burden (PAINS/BRENK/NIH),

\begin{equation}
    B = b_1(d_{charge}) + b_2(d_{interaction}) + b_3(d_{burden})
    \tag{F.8}
\end{equation}

\noindent This function is designed to penalise reactive linker environments that lead to irreversible payload-framework interactions. We also note that charge contributes to both safety and benign terms by design: safety captures acute electrostatic risk while benign score captures the payload interaction fit. Structural burden ($d_{burden}$) is estimated by calculating normalised count of reported warning matches for the fragments of the generated structure in either (i) PAINS (pan-assay interference); (ii) BRENK (reactive, toxic, or unstable fragments) or (iii) NIH (problematic functional groups based on NIH medicinal chemistry guidelines).\newline 
\indent \indent \indent The stability block (T) integrates metal stability, linker rigidity/symmetry, solubility-mediated dispersion behavior, and charge-release stability,
\begin{equation}
    T = t_1(d_{metal-stability}) + t_2 (d_{linker-stability}) + t_3 (d_{solubility}) + t_4 (d_{release})
    \tag{F.8}
\end{equation}

\noindent Solubility is recognised as important predictor of desirable bioavailability characteristics. Metal stability and safety are obtained from the curated dataset in \cite{menon2025guiding}, while the stability of the linker is determined based on rigidity and symmetry similar to the reward function for gas storage in \textbf{Supplementary Section F.1} above.\newline
\indent \indent \indent During method development, Bayesian optimisation was used to tune component and subcomponent weights under constrained known validation linker samples. The design objective combined toxicity-aware ranking constraints (to suppress high scores in severe toxicity classes), smooth score spread across viable chemotypes, and robustness across repeated sampled subsets. Final production weights were selected from the stable region identified across these runs and then fixed to preserve deterministic scoring as $\gamma _1 = 0.60$, $\gamma _2 = 0.14$, $\gamma _3 = 0.07$ and $\gamma _4 = 0.19$. Safety here is a rigid constraint and was assigned the highest weightage. Our calculated safety score qualitatively aligns well with expected trends (for instance, terephthalic acid $>$ ethylene glycol $>$ TCPP $>$ 2-methylimidazole $>$ Pyrene tetrabenzoic acid $>$ fentanyl).  \textbf{Supplementary Figure F.5} shows the distributions of the individual blocks and the distribution of the corresponding composite $R_{bio}$. All distributions remain broad and non-saturated, indicating that the function retains ranking resolution rather than collapsing near hard bounds.

\begin{figure}[h!]
\centering
\includegraphics[width = \textwidth]{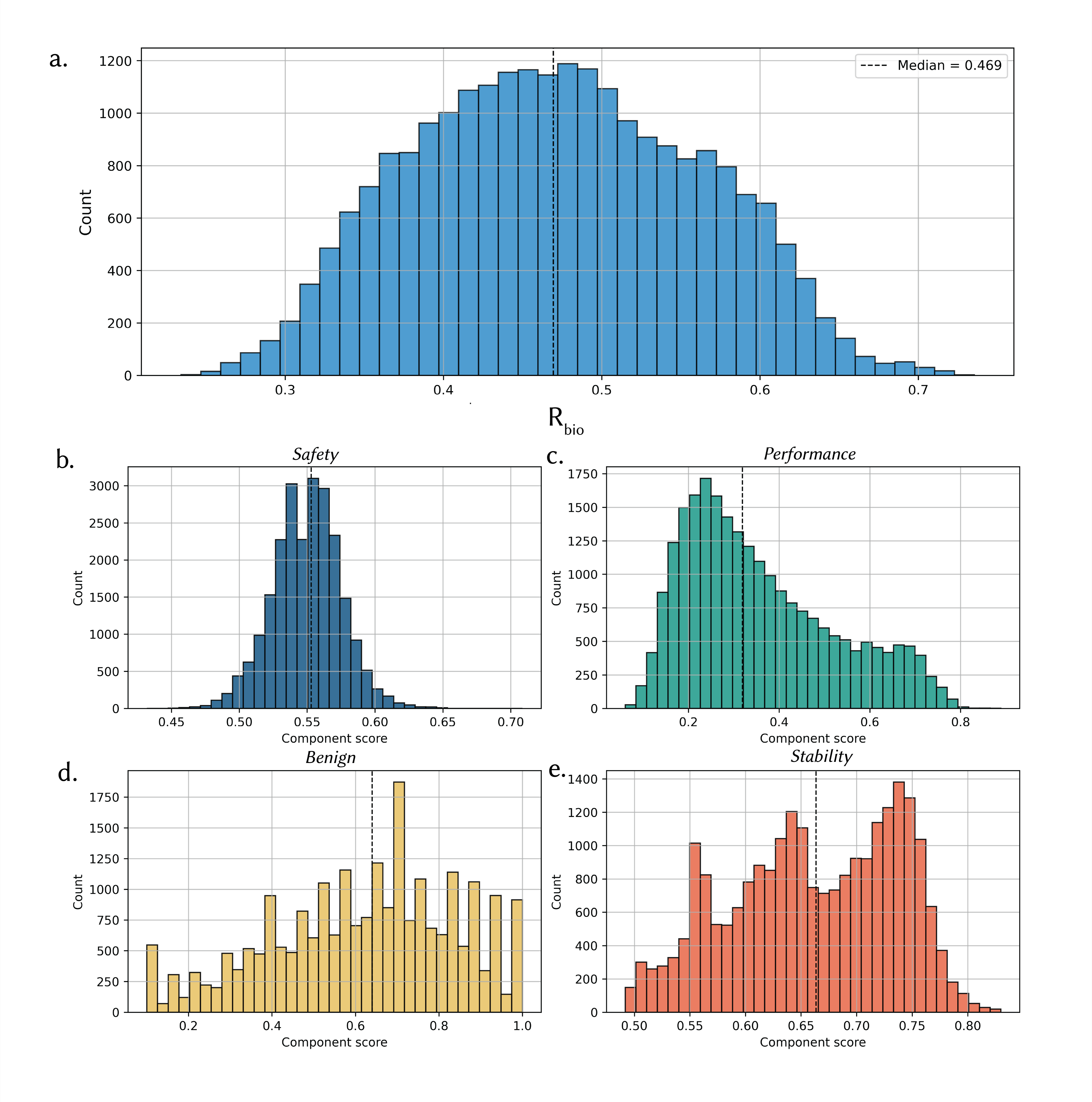}
\caption*{\textbf{Supplementary Figure F.5. a.} Distribution of the $R_{bio}$ scores of linkers from the training set (n = 25,000) of which 93.76\% produced valid scores after invalid SMILES cleaning/parsing. The total score distribution remains broad (with median = 0.469), indicating non-collapsed ranking capacity for molecular optimization. \textbf{b - e.} Distributions of individual components scores for \textbf{b.} safety (S); \textbf{c.} performance (P); \textbf{d.} benign-linker score (B) and \textbf{e.} stability (T). Each of these distributions is also non-collapsed to either extreme. We further impose rigid constraints during the design process, such as rejecting molecules with safety scores $< 0.50$.}
\end{figure}

%% file: sections/G.tex
Nexerra$^{R1}$ supports two distinct design modes for linker generation: (i) Direct design and (ii) Scaffold-constrained design. These design modes differ in how the latent variables are sampled and how geometric constraints are imposed. In practice, the direct design mode works best for 2-connected linkers and on nets with less strict topological constraints – for instance, \textit{fcu} and \textit{pcu} nets. On the other hand, the scaffold-constrained design mode works better for highly-connected linkers and nets with stricter topological constraints – such as \textit{csq}.\newline

\noindent \textbf{G.1. Direct design.} Here Nexerra$^{R1}$ generates linkers by sampling the local neighborhood around a user-specified seed molecule. Given a seed linker $x_{seed}$, the encoder approximates a posterior distribution,
\begin{equation}
    q_\phi (z|x_{seed}) = \mathcal{N}(z; \mu_{seed}, diag(\sigma^2_{seed}))
    \tag{G.1}
\end{equation}

\noindent where $\mu_{seed} = \mu_\phi (x_{seed})$ and $\sigma_{seed} = \sigma_\phi(x_{seed})$. To control the radius of the neighborhood, we sample embeddings from a scaled neighborhood of the posterior, i.e, 
\begin{equation}
    z = \mu_{seed} + \alpha\sigma_{seed} \cdot \epsilon, \ \epsilon \sim \mathcal{N}(0, I)
    \tag{G.2}
\end{equation}
\noindent where $\alpha > 0$ is a defined neighborhood radius. The smaller the $\alpha$, the more conservative the exploration while larger $\alpha$ values lead to more diverse molecules with respect to the seed. Each latent embedding is sampled using autoregressive decoding, 
\begin{equation}
    x \sim p_\theta (x|z)
    \tag{G.3}
\end{equation}
\noindent using temperature-scaling and nucleus sampling as described in \textbf{Supplementary Information Section B.4}. Generated molecules are then filtered first using strict chemical filters, followed by application-specific reward functions (see \textbf{Supplementary Information Section F}). The direct design mode thus enables rapid exploration of the local chemical neighborhood of seeds without imposing any scaffold-constraints.\newline   
\indent \indent \indent \textbf{Supplementary Figure G.1} shows the direct design method applied to several different classes of bidentate MOF linkers. These linkers post-generation are subjected to strict chemical filters to discard structures which are entirely implausible or highly synthetically complex. Here we show a subset of generated linkers with $R_{grav}$ scores $> 0.7 \times R_{grav}$ values of the seed.

\begin{figure}[h!]
\centering
\includegraphics[width = 0.7\textwidth]{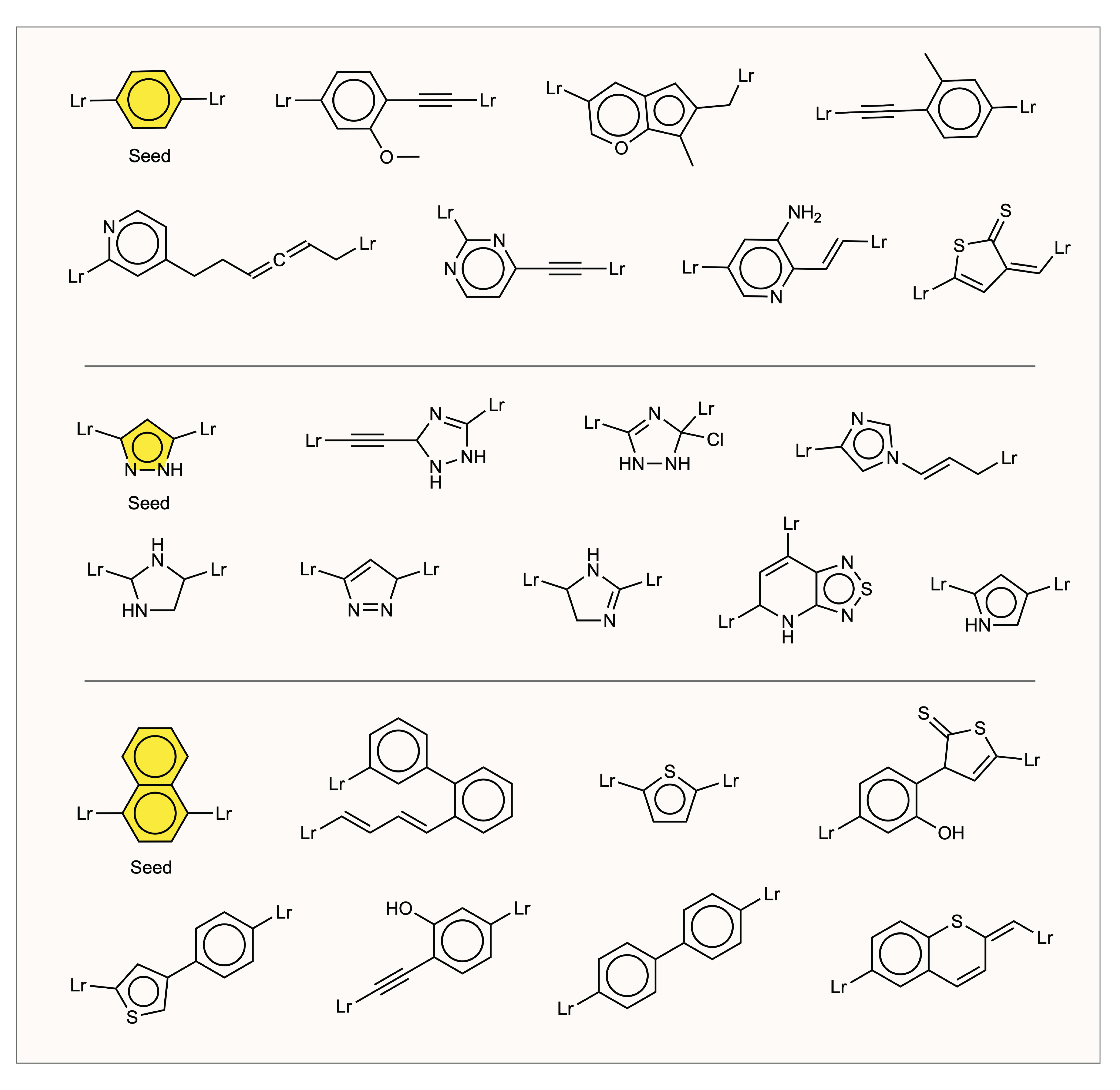}
\caption*{\textbf{Supplementary Figure G.1}. Direct design mode applied to different classes of bidentate MOF linkers. The seeds are highlighted in yellow.}
\end{figure} 

\noindent \textbf{G.2. Scaffold-constrained design.} Here the design is constrained around a pre-defined molecular core (or scaffold) that contains defined connecting handles (again denoted with the ‘[Lr]’ token for ease of processing). The goal of this mode is to preserve this scaffold – which is otherwise not possible in the direct design mode. Given a seed linker $x_{seed}$, we first decompose the linker as, 
\begin{equation}
    x_{seed} = (x_{scaff}, x_{arm})
    \tag{G.4}
\end{equation}

\indent \indent \indent The design method is agnostic to the number of arms, however, this number is preserved during the design process. In other words, this design mode does not change the connectivity of the input seed. After this decomposition, we apply the direct design method to $x_{arm}$ (see $G.1$ - $G.3$).\newline 
\indent \indent \indent Once new arms are generated, they are symmetrically grafted onto the core. First, we identify the anchor sites on the scaffold, denoted by the ‘[Lr]’ token. Here, the connection site is the atom directly bonded to ‘[Lr]’. Once these atoms are identified, the ‘[Lr]’ atoms are deleted from the scaffold. From the arm (which is ditopic), we delete one ‘[Lr]’ atom, while preserving the other one to indicate node-coordination moieties. We then attach this processed arm by forming a single bond between the scaffold connection site and the arm connection site. We then repeat this process for every connection site on the core. Finally, we perform a force-field optimization of the linker and carry out sanitization. As an illustrative example, \textbf{Supplementary Information Figure G.2} shows this design mode applied to the porphyrin-core.

\begin{figure}[h!]
\centering
\includegraphics[width = 0.8\textwidth]{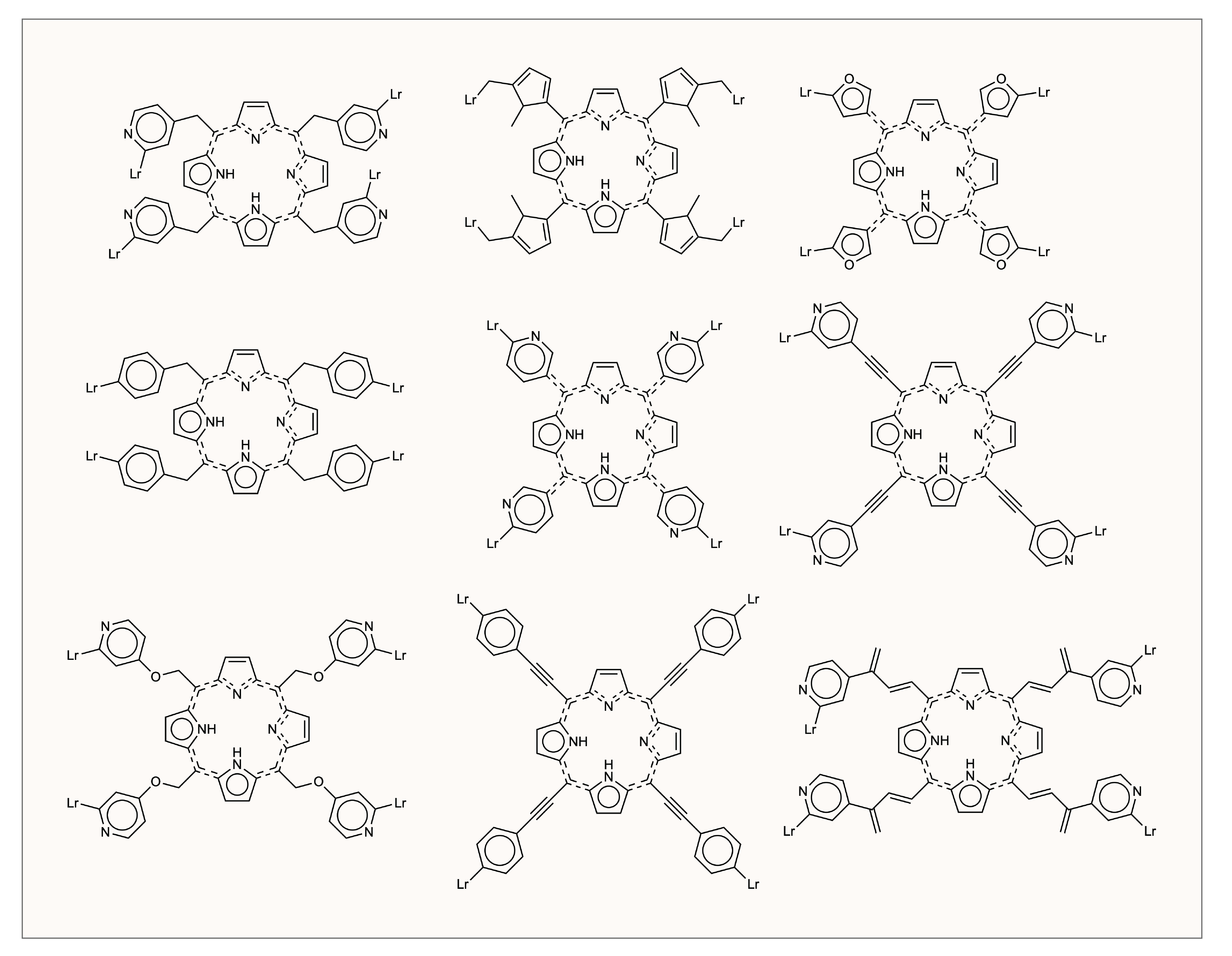}
\caption*{\textbf{Supplementary Figure G.2}. Scaffold-constrained design mode applied to the porphyrin-core.}
\end{figure}

%% file: sections/H.tex
\textbf{H.1. Synthetic procedure}\newline

\noindent \textbf{Synthesis of Linker (L).} The synthesis of L (\textbf{Supplementary Figure H.1}) follows a previously reported procedure \citep{matsunaga2011new}.\newline 

\begin{figure}[h!]
\centering
\includegraphics[width = 0.8\textwidth]{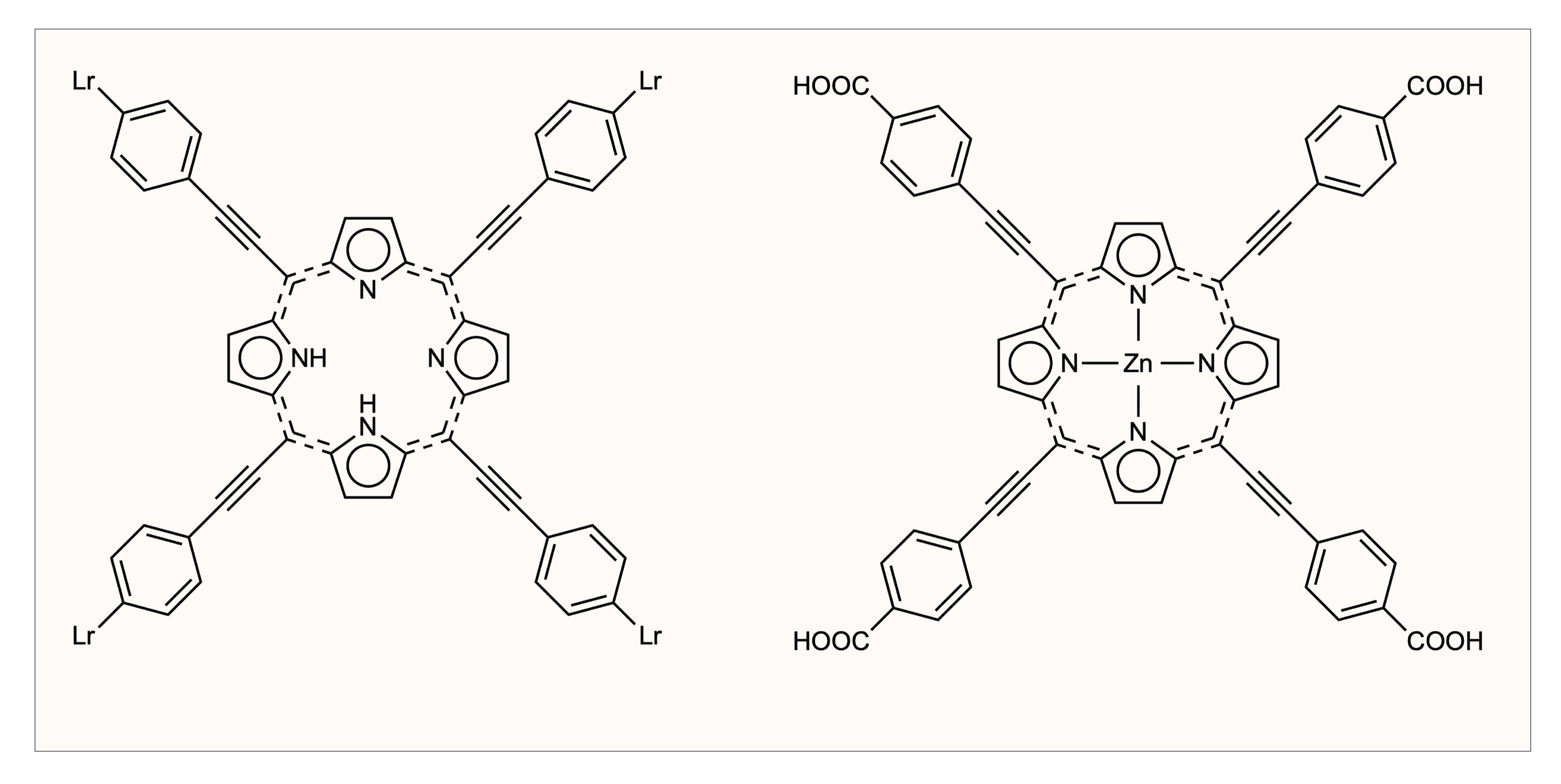}
\caption*{\textbf{Supplementary Figure H.1}. The porphyrin linker designed using Nexerra$^{R1}$ (left) and the linker (\textbf{L}) used for the synthesis of CU-525}
\end{figure}

\noindent\textbf{Synthesis of CU-525.} HfCl$_4$ (100 mg), H$_4$L (50 mg) and benzoic acid (1.25 g, 7.49 mmol) were dissolved in a mixture of DMA (25 mL) and HOAc (5 mL) through sonication. The resulting solution was sealed and heated in a 120 $^0$C oven for 48 h. Black cubic crystals were formed.\newline 

\noindent \textbf{H.2. Single Crystal X-ray Crystallography.} Single crystal X-ray diffraction experiments were performed on the PROXIMA 2A micro-focused beamline at Synchrotron SOLEIL ($\lambda$ = 0.729319 Å) using an EIGER X 9M 2D hybrid photon counting detector. X-ray data were collected at 100 K. Data integration and reduction were undertaken with Xia2 \citep{winter2010xia2}. No corrections for solvent scattering have been made. The structure was initially solved with SHELXT \citep{sheldrick2015shelxt} and refined by full-matrix least squares refinement on F$^{2}$ by the SHELXTL-2014 \citep{sheldrick2015crystal} using the Olex2-1.5 \citep{dolomanov2009olex2} software package. Standard crystallographic methods were used where necessary, including constraints, restraints, and rigid bodies. All carbon-bound hydrogen atoms were added in idealized positions and refined using a riding model. Restraints were used to obtain reasonable parameters. H-atoms were refined isotropically, while the other atoms were refined anisotropically. Crystal data and details of the X-ray data collection are given in \textbf{Supplementary Table H.1}.\newline 

\begin{table}[t!]
  \caption*{\textbf{Supplementary Table H.1}. Crystal data and structure refinement for CU-525}
  \centering
  \begin{tblr}{
      colspec={ll},
      row{1}={font=\bfseries},
      column{1}={font=\itshape},
      row{even}={bg=gray!10},
    }
              & CU-525 (CCDC: 2514566) \\
    \toprule
    Formula & C$_{38.67}$H$_{54.67}$HfN$_{7.33}$O$_{5.33}$Zn$_{0.5}$  \\
    F.W. (g mol$^{-1}$) & 918.73 \\
    Temperature/K &	 100 \\
    Wavelength/\AA  & Synchrotron ($\lambda = 0.729319$) \\
    Crystal system & Cubic \\
    Space group &  Pm-3m \\
    $a = b = c$ / \AA & 22.7093(2) \\
    $\alpha = \beta = \gamma$ / $^\circ$ & 90 \\
    Volume / \AA$^3$, $Z$ & 11711.5(3), 6 \\
    $\rho_{calcg}$/cm$^3$ &	0.782 \\
    $\mu$/mm$^{-1}$ &	1.608 \\
    F(000) & 2806.0 \\
    $2\theta$ range for data collection / $^\circ$ & 2.602 to 54.206 \\
    Reflections collected &	117300 \\
    Independent reflections & 2228 [$R_{\mathrm{int}} = 0.1933$, $R_{\sigma} = 0.0858$] \\
    Data/restraints/parameters & 2228/110/68 \\
    Goof & 0.894 \\
    Final $R$ indexes [$I \ge 2\sigma(I)$] & $R_1 = 0.0928$, $wR_2 = 0.2712$ \\
    Final $R$ indexes [all data] & $R_1 = 0.1193$, $wR_2 = 0.2811$ \\
    Largest diff. peak/hole / e \AA$^{-3}$ & 2.60/-2.48 \\
    \bottomrule
  \end{tblr}
\end{table}

\noindent \textbf{H.3. Comparison of structural models.} \textbf{Supplementary Figure H.2} compares the structural model obtained from the topological assembly of the Nexerra$^{R1}$-generated linker ($\Phi$- 11) with the experimentally resolved single-crystal structure of CU-525. The model structure is assembled using the Zr$_6$-oxo cluster (consistent with MOF-525), while CU-525 is synthesized experimentally using an Hf$_6$-oxo cluster. Due to nearly identical coordination geometry, this substitution is not expected to alter the underlying net. At a global topology and connectivity level, $\Phi$-11 reproduces the coordination environment and node connectivity of CU-525; however, local differences are observed in some linker arms. These arise because the assembly algorithm selects an alternate but symmetry-equivalent orientation – essentially ‘flipping’ the placement of the arm. This occurs because the arm placement is geometrically invariant and the algorithm is unable to discriminate between the alternate placement conditions. Resolving this would require substantial refinement of the assembly algorithm itself and lies outside the scope of the present work.\newline
\indent \indent \indent Simulated powder X-ray diffraction (PXRD) patterns of $\Phi$-11 and CU-525 are also shown in \textbf{Supplementary Figure H.2}. Both patterns are in close agreement with each other, with no additional peaks observed in the generated structure. We, however, observe a slight systematic shift in the peak positions of the generated structure relative to CU-525 which arise as an effect of structural relaxation. There are differences in relative peak positions as well, which are attributed to variations in either atomic positions or linker conformations. Overall, these observations suggest structural equivalence with minor conformational deviations. 

\begin{figure}[h!]
\centering
\includegraphics[width = 0.8\textwidth]{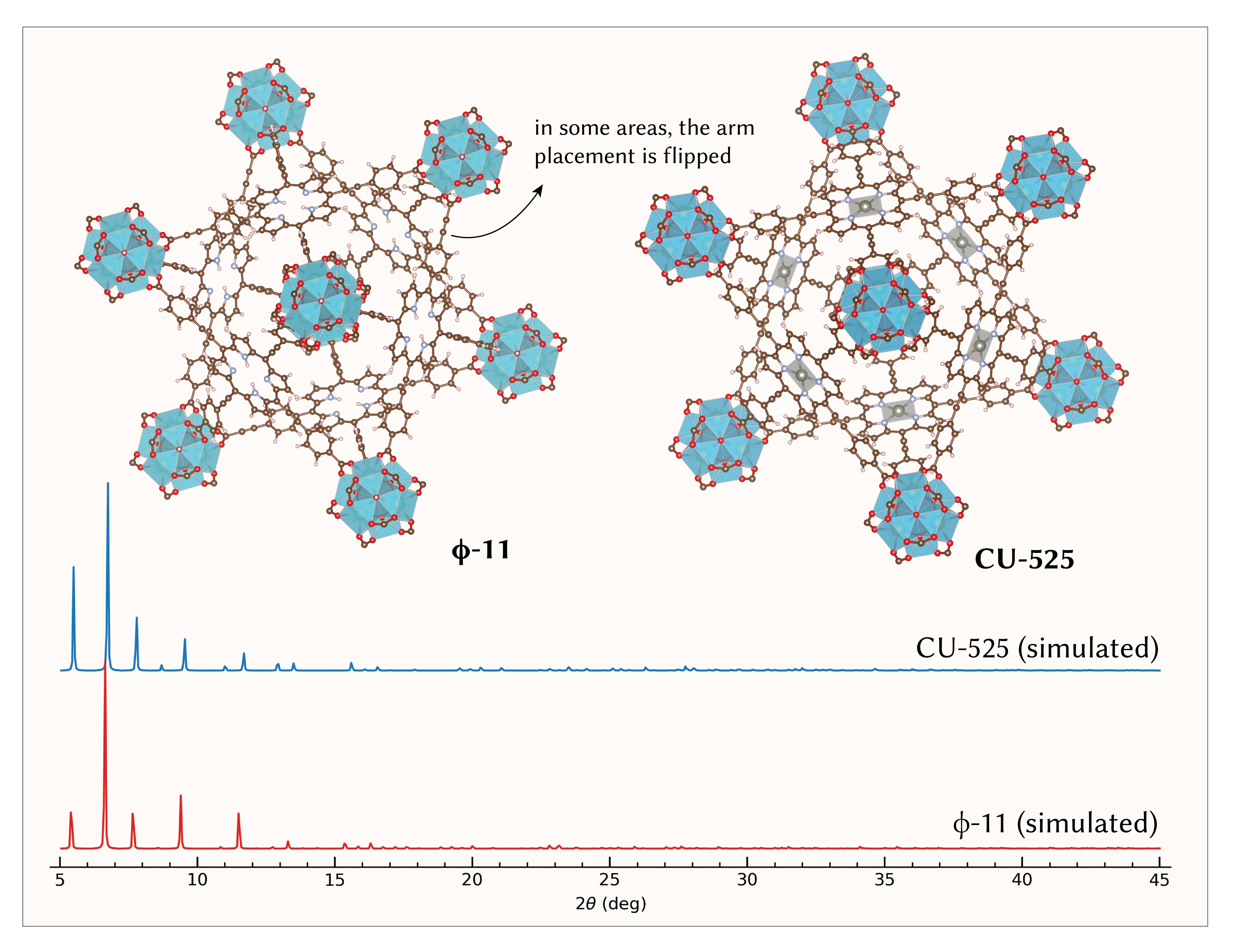}
\caption*{\textbf{Supplementary Figure H.2}. (Top) Optimized structural model of $\Phi$-11 (left) compared to the structure of CU-525 resolved using SCXRD as described above. We note that the MOF assembly algorithm flips the placement of the arm in certain instances leading to conformation deviations from the experimentally resolved structure. (Bottom) Simulated PXRD patterns of $\Phi$-11 (red) and CU-525 (blue). Please note that the patterns have been max-normalized.}
\end{figure} 

%% file: sections/I.tex
\textbf{I.1. Flow matching formulation.} We use the flow-matching formulation introduced by \cite{lipman2022flow} and extended by \cite{tong2023improving}. Below, we provide only the essential mathematical definitions and refer readers to the original and foundational references for full derivations. Starting with the latent space learnt by Nexerra$^{R1}$, let:
\begin{equation}
 {(z_i, y_i)}_{(i=1)}^N,\ z_i \in \mathbb{R}^D, y_i \in \mathbb{R}
 \tag{I.1}
\end{equation}
\noindent denote the latent encodings $z_i$ of the training dataset and $y_i$ the associated scalar property (e.g. anchor – anchor span, logP, or, in principle, any reward function). The latent variables are then standardised as: 

\begin{equation}
 \mu_z = \frac{1}{N}\sum_{i = 1}^Nz_i, \sigma _z = \sqrt{\left(\frac{1}{N}\sum_{i = 1}^N(z_i - \mu _z)\odot (z_i - \mu _z) + \epsilon \right)}
 \tag{I.2}
\end{equation}
\noindent and we define:

\begin{equation}
    \tilde{z}_i= \frac{(z_i - \mu_z)}{\sigma_z} 
 \tag{I.3}
\end{equation}

\noindent The flow model and associated losses are defined in this standardised latent space. This is then inverted during inference. For training, we replace the standard Gaussian prior with an empirical latent prior defined over a finite set of encoder embeddings. Initial latent states are sampled uniformly from a percentile-filtered subset of this latent bank. We define a percentile-based threshold ($\tau_s$) on $\{y_i\}$ such that for sorted property values $y_1 \geq y_2 \geq ... \geq y_N$, we have a linear percentile scheduler:

\begin{equation}
    P_s = P_{start} + (P_{end} - P_{start}) \times \frac{n}{S}
    \tag{I.4}
\end{equation}

\noindent where $n$ is the current training step, and $S$ is the total number of training steps. For maximising a property (e.g. length) $P_{end} > P_{start}$ while for minimizing a property (e.g. logP) $P_{start} > P_{end}$. Given $P_n \in [0,100]$, we select an index $k_n = [N\frac{(100-P_n)}{100}]$ and define the current threshold as:

\begin{equation}
    \tau_s = Y_{k_n}
    \tag{I.5}
\end{equation}

\noindent The conditional target distribution at step $n$ is the empirical distribution over latents whose property exceeds the threshold ($\tau_n$):

\begin{equation}
    q_{\tau_s}(z) \propto \sum_{i=1}^N \{y_i \geq \tau_n \}\delta (z - \tilde{z}_i)
    \tag{I.6}
\end{equation}

\noindent For a contrastive unconditional sampling (used for classifier-free guidance, discussed shortly) we define a larger subset of the latent distribution (OR the full distribution) as:

\begin{equation}
    q_{uncond}(z) \propto \sum _{i = 1} ^N \delta(z - \tilde{z}_i)
    \tag{I.7}
\end{equation}

\noindent We parameterize a continuous normalizing flow (CNF) in the latent space by an ordinary differential equation (ODE):

\begin{equation}
    \frac{d(z)}{dt} = f_\theta (t, z(t), c), \ t \in [0, 1], \  z(0) \sim p_0
    \tag{I.8}
\end{equation}

\noindent where $z(t) \in \mathbb{R}^D$ is the latent state at time $t$, $c \in \mathbb{R}$ is the scalar conditioning variable and $f_\theta:[0,1] \times \mathbb{R}^D \times \mathbb{R} \rightarrow \mathbb{R}^D$ is the neural net vector field. We implement this vector field as a neural network $NN$ (defined in \textbf{Supplementary Information Section I.2}) operating on the concatenation of the latent ($z$), time embedding ($t$) and condition ($c$):

\begin{equation}
    f_\theta (z, t, c) = NN([z, \gamma(t), c])
    \tag{I.9}
\end{equation}

\noindent where $\gamma(t) \in \mathbb{R}^{2d_{Fourier}}$ is a Fourier feature embedding of time – $\gamma(t) = (sin(2\pi Bt),cos(2\pi Bt))$ – similar to the positional encoding performed on the training data before passing it to the transformer encoder. $B \in \mathbb{R}^{d_{Fourier}}$ is a random frequency vector ($d=64$). For a fixed condition $c$, integrating this ODE from $t=0\rightarrow t = 1$ defines a flow map:

\begin{equation}
    \Phi_\theta^c : \mathbb{R}^D \rightarrow \mathbb{R}^D,\ z_1 = \Phi_\theta ^c(z_0)
    \tag{I.10}
\end{equation}

\noindent This flow map approximates the optimal transport between the empirical prior ($p_0$) towards the conditional empirical distribution $q_\tau(z)$.\newline 

\indent \indent \indent We train $f_\theta$ using the exact optimal transport-conditional flow matching (OT-CFM) between the empirical prior and the conditional (OR unconditional) empirical latent distribution \citep{tong2023improving}. For a training sample ($s$), we sample the base latents: $z_0 \sim p_0$, conditional target: $z_1^{cond} \sim q_{\tau_s}$ and unconditional target: $z_1^{uncond} \sim q_{uncond}$. These are paired with corresponding conditions: $c^{cond} = \tau_s, c^{uncond} = c_{uncond} \leftarrow $ (in our case $\varnothing = 0.0$). In order to train both conditional and unconditional flows in a single network, we apply conditional dropout with probability $p_{drop}$. For each sample in the mini-batch, we choose:

\begin{equation}
    P\left((z_1, c) = (z_1^{uncond}, c_{uncond})\right) = p_{drop},
    \tag{I.11.1}
\end{equation}
\begin{equation}
    P\left((z_1, c) = (z_1^{cond}, c_{cond})\right) = 1 - p_{drop}
    \tag{I.11.2}
\end{equation}

\noindent For each pair $(z_0,z_1)$, we define the displacement interpolation (the OT geodesic in Euclidean space):

\begin{equation}
    x_t = (1 - t)z_0 + tz_1, \ t \in [0, 1]
    \tag{I.12}
\end{equation}

\noindent Under the quadratic cost OT formulation, the exact conditional OT flow along this path has a constant velocity:

\begin{equation}
    v^*(z_0, z_1) = z_1 - z_0 \leftarrow t \ independent 
    \tag{I.13}
\end{equation}

\noindent OT-CFM leverages this formulation to construct a simple closed-form target vector field. The training objective minimizes the discrepancy between the model vector field and this OT-flow at random intermediate times:

\begin{equation}
    \mathcal{L}_{OT-CFM}(\theta) = \mathbb{E}_{z_0 \sim p_0, (z_1, c) \sim q_s, t \sim \mathcal{U}(0, 1)}||f_{\theta}(t, x_t, c) - (z_1 - z_0)||_2^2
    \tag{I.14}
\end{equation}

\noindent where $q_s$ is a mixture over conditional and unconditional latent samples. The training is performed in mini-batches and $t$ is sampled \textit{i.i.d} from $\mathcal{U}(0,1)$ per example.\newline 
\indent \indent \indent During sampling, we use classifier-free guidance (CFG) \citep{ho2022classifier} to bias the flow model towards high-threshold conditions without having to train a separate classifier. Given a desired threshold $c^*=\tau^*$ and a CFG scale – $m \geq 0$, we define conditional and unconditional vector fields at a point ($t,z$) as:

\begin{equation}
    v_{uncond}(t, z) = f_{\theta}(t, z, c_{uncond}), v_{cond}(t, z; c^*) = f_\theta(t, z, c^*)
    \tag{I.15}
\end{equation}

\noindent The guided vector field is then:
\begin{equation}
    v_{CFG}(t, z; c^*, m) = v_{uncond}(t, z) + m(v_{cond}(t, z; c^*) - v_{uncond}(t, z))
    \tag{I.16}
\end{equation}

\noindent We generate samples by solving the guided ODE:
\begin{equation}
    \frac{dz(t)}{dt} = v_{CFG}(t, z(t); c^*, s),\ z(0) \sim p_0
    \tag{I.17}
\end{equation}

\noindent from $t = 0 \rightarrow t = 1$. The ODE is integrated with an adaptive Dormand-Prince solver. For $m = 0$, this is reduced to an unconditioned model and for $m = 1$, this is the plain conditional model. $m > 1$ amplifies condition-dependent output.\newline 

\noindent \textbf{I.2. Neural network architecture.} The core network is a stack of N residual blocks (ResBlocks) \citep{he2016deep}. For our implementation, we use $N = 8$. Each block uses FiLM conditioning \citep{perez2018film}. Given a block input $x \in \mathbb{R}^H$, we first apply a LayerNorm ($LN$):

\begin{equation}
    \tilde{x} = LN(x)
    \tag{I.18}
\end{equation}

\noindent A FiLM module produces per-feature scale and shift parameters $(\gamma, \beta) \in \mathbb{R}^H$ from the conditioning vector:

\begin{equation}    
    (\gamma, \beta) = FiLM(h_c),\ FiLM: \mathbb{R}^H \rightarrow \mathbb{R}^{2H}
    \tag{I.19}
\end{equation}
\noindent where $h_c$ is a hidden conditioning vector. The normalized activations are then modulated per-feature as,

\begin{equation}
    \hat{x} = \tilde{x} \odot (1 + \gamma) + \beta
    \tag{I.20}
\end{equation}

\noindent These are then passed through a MLP with SiLU activations. The FiLM layer is intended to improve the stability of the deep residual stacks. Following this, we add $LN$ and linear projection back to the latent dimension.\newline 

\noindent \textbf{I.3. Training and evaluation.} Nexerra$^{R1}$-Flow was trained over 200,000 steps with evaluation every 20,000 steps. At each evaluation, 5000 samples were generated and shift in the property distribution and transport fidelity was logged. \textbf{Supplementary Figure I.1} shows the corresponding training logs. The transport accuracy was quantified via relative RMSE (or rRMSE) and mean cosine similarity between predicted and reference latent displacements. Across the training steps, rRMSE increased moderately from $\sim$ 0.02 to $\sim$ 0.035 ($\sim$ 0.030 at final checkpoint), while cosine similarity decreased gradually from $\sim$ 0.97 to $\sim$ 0.92 – 0.94. These trends are reflective of progressively stricter percentile scheduler – i.e., the target slice moving toward higher percentiles. As the target to meet gets stricter, the required displacement increases in magnitude. Importantly, cosine similarity remained high ($>$ 0.92) throughout training, indicating strong directional alignment of the learned vector field with the reference OT displacement.\newline
\indent \indent \indent Sampled molecules exhibited a systematic upward shift in the target property distribution (here the anchor – anchor span) over training. The median span increased, while the interquartile range remained stable across checkpoints. This indicates redistribution of probability mass toward higher-percentile regimes without collapse. Notably, at step 180,000 we recorded the highest relatively median anchor-anchor length of 9.05 Å. We thus, selected this checkpoint for further production. Overall, training dynamics are stable and consistent with successful optimal-transport-based conditional generation.\newline 

\begin{figure}[t!]
\centering
\includegraphics[width = \textwidth]{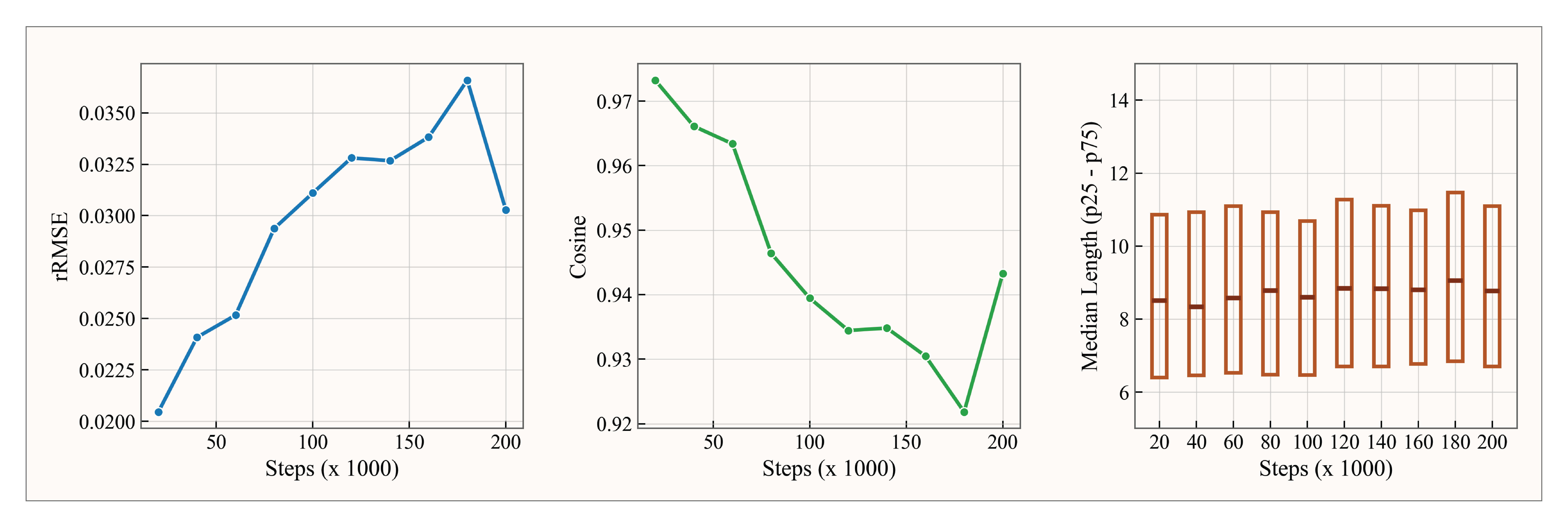}
\caption*{\textbf{Supplementary Figure I.1.}(Left to right). rRMSE of predicted and target latent displacement; mean-cosine similarity between the predicted and target latent displacements; median anchor – anchor span for n = 5000 sampled molecules from the model checkpoint. Please note: during training the target is monotonically increased from the 70$^{th}$ percentile to the 95$^{th}$ percentile – i.e., the training gets progressively harder as the steps increase, which reflects in the growing rRMSE and dropping cosine.  }
\end{figure} 

\begin{figure}[t!]
\centering
\includegraphics[width = 0.8\textwidth]{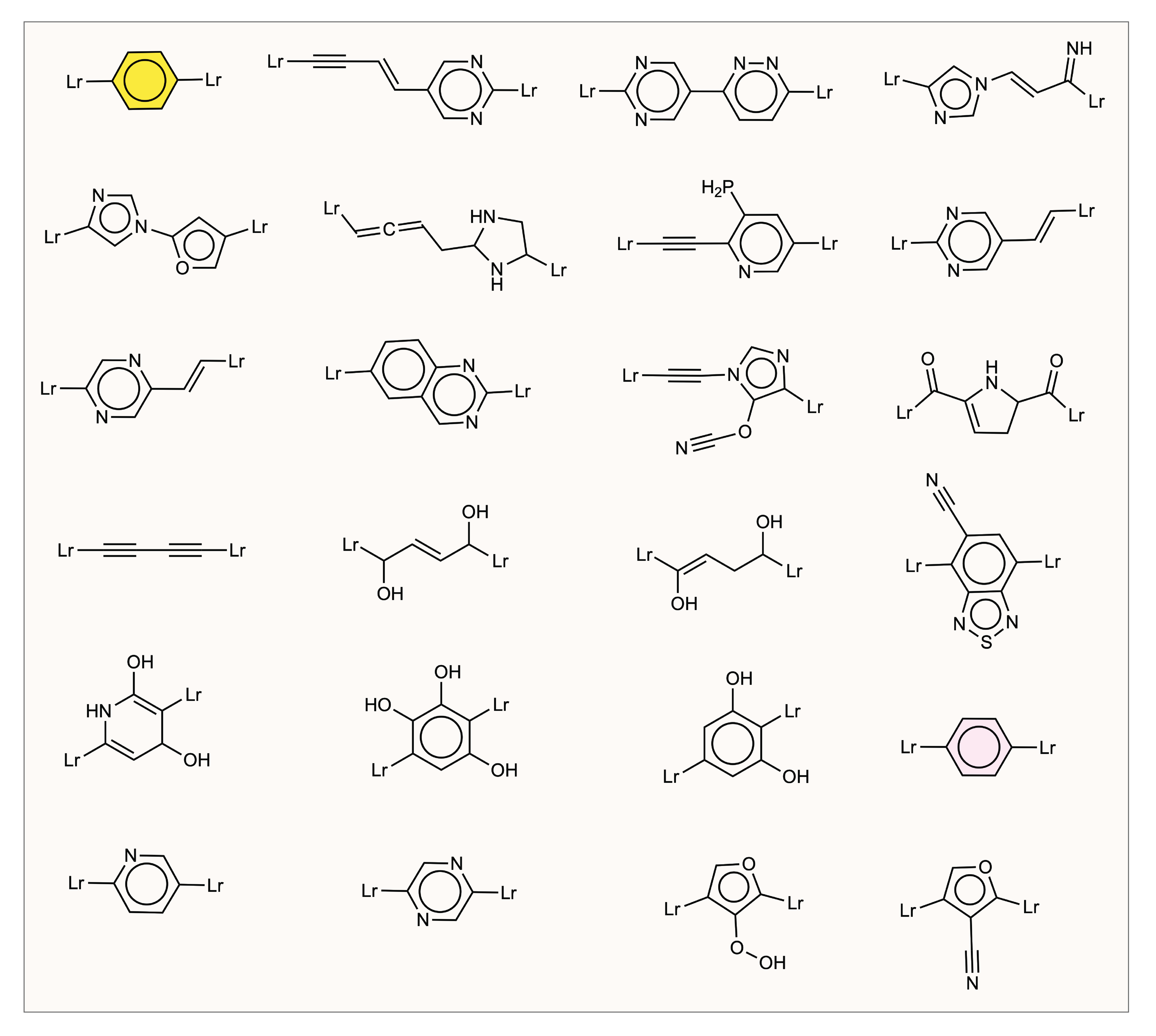}
\caption*{\textbf{Supplementary Figure I.2.} A subset of linkers generated by Nexerra$^{R1}$-Flow using a BDC seed (yellow). These linkers were selected post-filtering for geometric compatibility with the \textit{fcu} net. The linker highlighted in pink is a rediscovery of the input seed.}
\end{figure}

\noindent \textbf{I.4. Flow generated linkers and structures.} We shift our focus to linkers and their corresponding structures generated using the trained flow model. Each optimized structure has been appended to the data supporting this manuscript. Starting with the BDC linker as a seed, \textbf{Supplementary Figure I.2} shows sample linkers generated by Nexerra$^{R1}$-Flow subject to strict geometric filters to ensure compatibility with the target \textit{fcu} net. The main filter implemented here was restricting selection to those linkers whose anchor – anchor angle $>$ 160$^0$ – i.e., a linearity constraint. 

\indent \indent \indent \textbf{Supplementary Figure I.3} shows optimized structures generated using these linkers on an \textit{fcu} net using a Zr$_6$-oxo cluster. While in most cases structural assembly was successful, we observed instances where assembly fails – most commonly, due to overlapping atoms, or additionally coordinating moieties close to the placed handles. Success rates however depend on topological and geometric complexity of the target net.

\begin{figure}[t!]
\centering
\includegraphics[width = 0.8\textwidth]{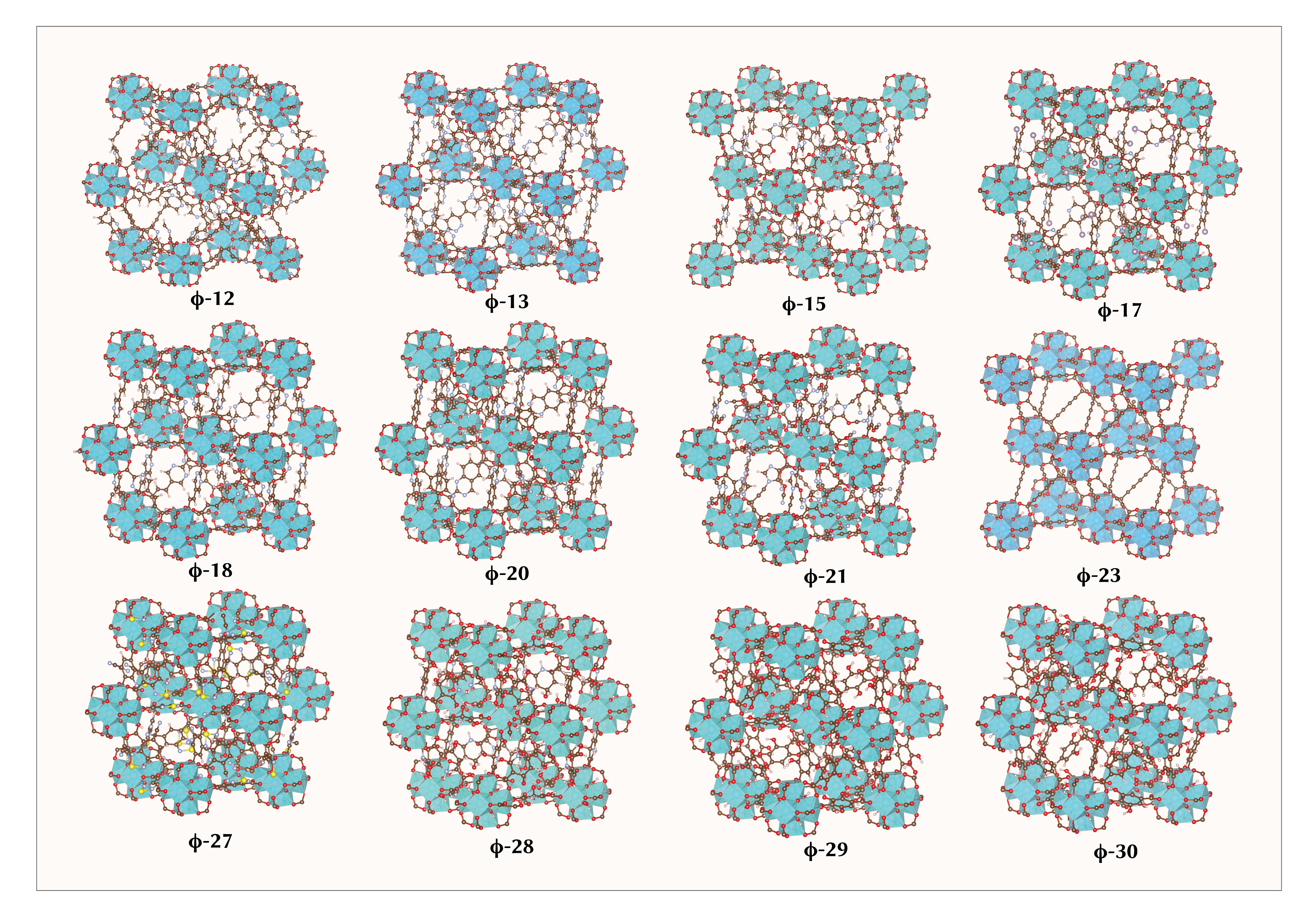}
\caption*{\textbf{Supplementary Figure I.3.} A subset of optimized structures assembled on the \textit{fcu} net from linkers generated from Nexerra$^{R1}$-Flow.}
\end{figure} 

%% file: sections/J.tex
Gravimetric and volumetric methane uptake capacities were obtained from grand canonical Monte Carlo (GCMC) simulations, in which chemical potential, volume, and temperature were held constant. Simulations were performed with the RASPA code \citep{dubbeldam2016raspa}, treating both framework and adsorbate molecules as rigid. During each Monte Carlo cycle, adsorbates were allowed to undergo translation, insertion, and swap moves – each with equal probability. Each simulation comprised 30,000 cycles; the first 15,000 cycles were treated as equilibration, following which ensemble averages were collected over 15,000 production cycles. The number of trial moves per cycle was either the current number of adsorbates in the system or 20, whichever is the largest. 
Adsorbate-framework and adsorbate-adsorbate interactions were described using Lennard-Jones (LJ) potentials combined with electrostatic interactions. LJ interactions were truncated and shifted to zero at 12.5 Å, and Ewald summation was applied for long-range electrostatics. LJ parameters for CH$_4$ were adopted from \citep{martin2001effect}. Framework LJ parameters were chosen from the Dreiding force field \citep{mayo1990dreiding}, where available, and with UFF parameters otherwise \citep{rappe1992uff}. No framework partial charges were assigned since the CH$_4$ model adopted does not include partial charges. For interactions lacking explicit parametrization, cross-interactions were determined using Lorentz-Berthelot mixing rules. We append a set of sample GCMC scripts in the data that supports this manuscript for readers who wish to reproduce these calculations.  